\documentclass[lettersize,journal]{IEEEtran}
\usepackage{amsmath,amsfonts,amssymb}
\usepackage{array}
\usepackage{pifont}
\usepackage[caption=false,font=normalsize,labelfont=sf,textfont=sf]{subfig}
\usepackage{textcomp}
\usepackage{stfloats}
\usepackage{url}
\usepackage{verbatim}
\usepackage{graphicx}
\usepackage{cite}
\usepackage[normalem]{ulem}
\usepackage{mathtools}
\usepackage{multirow,multicol}
\usepackage{lineno,hyperref}
\usepackage{booktabs}
\usepackage[table]{xcolor}
\usepackage[ruled,vlined,linesnumbered]{algorithm2e}
\begin{document}

\title{Diffusion-Guided Adversarial Perturbation Injection for Generalizable Defense Against Facial Manipulations}

\author{Yue Li,~\IEEEmembership{Member, IEEE}, Linying Xue, Kaiqing Lin, Hanyu Quan,~\IEEEmembership{Senior Member, IEEE}, \\Dongdong Lin$^{\ast}$,~\IEEEmembership{Member, IEEE}, Hui Tian$^{\ast}$,~\IEEEmembership{Senior Member, IEEE}, Hongxia Wang,~\IEEEmembership{Member, IEEE}, Bin Wang
        % <-this % stops a space
\thanks{Yue Li, Linying Xue, Hanyu Quan, Dongdong Lin and Hui Tian are with the College of Computer Science and Technology, National Huaqiao University, Xiamen 361021, China, and also with the Xiamen Key Laboratory of Data Security and Blockchain Technology, Xiamen 361021, China (e-mail:  liyue\_0119@hqu.edu.cn; 23014083061@stu.hqu.edu.cn; quanhanyu@gmail.com; dongdonglin8@gmail.com; htian@hqu.edu.cn).}% <-this % stops a space
\thanks{Kaiqing Lin is with Shenzhen Key Laboratory of Media Security, Shenzhen University, Shenzhen 518060, China (email: linkaiqing2021@email. szu.edu.cn)}
\thanks{Hongxia Wang is with the School of Cyber Science and Engineering,
Sichuan University, Chengdu 610207, China (e-mail: hxwang@scu.edu.cn)}
\thanks{Bin Wang is with Zhejiang Key Laboratory of Artificial Intelligence of Things (AIoT) Network and Data Security, Hangzhou 310053, China (e-mail: wangbin02@xidian.edu.cn)}
\thanks{\textit{Corresponding author: Dongdong Lin and Hui Tian}}
}

\markboth{Journal of \LaTeX\ Class Files,~Vol.~14, No.~8, August~2021}%
{Shell \MakeLowercase{\textit{et al.}}: A Sample Article Using IEEEtran.cls for IEEE Journals}


\maketitle

\begin{abstract}
Recent advances in GAN and diffusion models have significantly improved the realism and controllability of facial deepfake manipulation, raising serious concerns regarding privacy, security, and identity misuse. Proactive defenses attempt to counter this threat by injecting adversarial perturbations into images before manipulation takes place.
However, existing approaches remain limited in effectiveness due to suboptimal perturbation injection strategies and are typically designed under white-box assumptions, targeting only simple GAN-based attribute editing. These constraints hinder their applicability in practical real-world scenarios.
In this paper, we propose AEGIS, the first \textit{diffusion-guided} paradigm in which the \uline{A}dv\uline{E}rsarial facial images are \uline{G}enerated for \uline{I}dentity \uline{S}hielding. 
We observe that the limited defense capability of existing approaches stems from the \textit{peak-clipping constraint}, where perturbations are forcibly truncated due to a fixed $L_\infty$-bounded. 
To overcome this limitation, instead of directly modifying pixels, AEGIS injects adversarial perturbations into the latent space along the DDIM denoising trajectory, thereby decoupling the perturbation magnitude from pixel-level constraints and allowing perturbations to adaptively amplify where most effective.
Leveraging a pre-trained DDIM model without any additional training, the intrinsic noise-refinement mechanism of AEGIS naturally balances the strength and imperceptibility of perturbations.
The extensible design of AEGIS allows the defense to be expanded from purely white-box use to also support black-box scenarios through a gradient-estimation strategy.
Extensive experiments across GAN and diffusion-based deepfake generators show that AEGIS consistently delivers strong defense effectiveness while maintaining high perceptual quality.
In white-box settings, it achieves robust manipulation disruption, whereas in black-box settings, it demonstrates strong cross-model transferability. Moreover, the generated adversarial images of AEGIS retain minimal identity-recognizable information, mitigating facial stigmatization and reducing identity-misuse risks.
\end{abstract}


\section{Introduction}
\IEEEPARstart{F}{acial} deepfake refers to the use of deep generative models to synthesize or manipulate human faces in images or videos with highly realistic visual fidelity \cite{deepfake_survey}. Since its emergence, deepfake has rapidly evolved into a pervasive research and societal topic. The rapid progress of generative architectures, particularly GANs~\cite{goodfellow2014gan} and diffusion models~\cite{DDIM,DDPM}, has significantly lowered the barrier to producing convincing facial forgeries.
While deepfake techniques enable legitimate applications such as film post-production and digital avatar creation, their misuse has caused severe security and ethical concerns. Non-consensual explicit content, identity impersonation, and political misinformation have already been reported, demonstrating deepfake's ability to erode personal privacy and public trust~\cite{maras2019determining,ajder2019state,westerlund2019emergence,zhu2025hiding}.

\begin{figure}[t]
    \centering
    \includegraphics[width=1\linewidth]{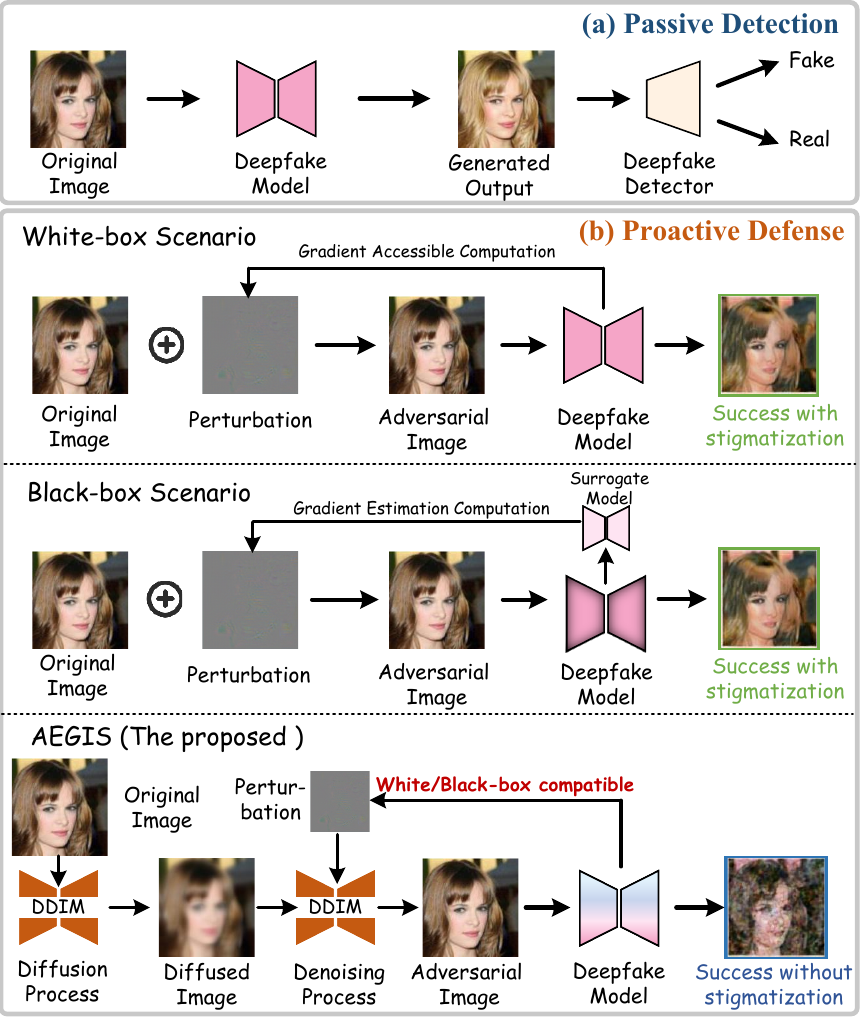}
    \vspace{-0.5cm}
    \caption{Defense taxonomy. Passive detection identifies deepfakes post-generation, whereas proactive defense prevents manipulation. AEGIS is the first method compatible with both white-box and black-box proactive defense settings.}
    \vspace{-0.7cm}
    \label{fig:abstract}
\end{figure}

Existing research on deepfake security has predominantly focused on passive detection (Fig.~\ref{fig:abstract}(a)), where models are trained to distinguish manipulated content from authentic media \cite{wang2024deepfake,lu2023detection,yin2023dynamic,wang2025idcnet,deng2024towards,zhang2024deepfake}. Despite progress, detection remains fundamentally reactive: it only takes effect after forged content has been generated and potentially disseminated. Consequently, it cannot prevent the initial spread of malicious deepfakes nor mitigate the immediate psychological, reputational, or social harm inflicted on victims. Moreover, the arms race between detectors and deepfake generators creates an unintended feedback loop—every improvement in detection models implicitly drives deepfake techniques toward greater realism and evasiveness \cite{mittal2020emotions,Lin_2024_CVPR}. Compounding the issue, recent studies show that deepfake detectors are vulnerable to adversarial manipulation; small, imperceptible perturbations can induce misclassification, leading to false negatives and enabling forged content to bypass detection~\cite{carlini2020evading,hussain2021adversarial,meng2024ava}. These limitations highlight that passive detection alone is insufficient, motivating a shift toward proactive defense strategies that disrupt deepfake generation at the source.

Hence, a growing body of research has shifted toward proactive defense, where adversarial perturbations are injected into benign facial images to disrupt deepfake generation. These methods can be grouped according to whether the deepfake model’s architecture or gradients are available (Fig.~\ref{fig:abstract}(b)): white-box approaches assume full access\cite{2020_white_blur,ijcai2022p107,2024_saliency_aware,2024_union-saliency,2024_DF_RAP}, whereas black-box approaches rely on model-agnostic perturbation transfer\cite{huang2021initiative,2023_TIFS_Black_box,zhang2025robust}. Regardless of the setting, the core challenge reduces to navigating a three-way trade-off: maximizing defense success rate while maintaining perturbation imperceptibility, and balancing a third objective—robustness for white-box defenses or transferability for black-box defenses. Strengthening one objective often degrades another, meaning these methods are forced to make compromises along this triangle.

To preserve visual imperceptibility, prior methods constrain perturbation magnitude via a fixed $L_\infty$ bound. Through our analysis, we reveal that this optimization regime introduces a previously overlooked \textbf{peak-clipping constraint}: \textit{perturbation peaks that are most effective in disrupting manipulation are forcibly clipped to satisfy the visual constraint, resulting in uniform and suboptimal perturbation distributions} (see Sec.~\ref{sec.motivation} for detailed analysis). The suppressed perturbation strength further causes adversarial facial images to retain recognizable identity cues from the original face, which not only weakens defense effectiveness but also raises the risk of residual identity exposure, leaving the victim vulnerable to facial stigmatization and secondary misuse. Consequently, current defenses struggle to simultaneously achieve strong manipulation disruption and high perceptual quality.

While the peak-clipping constraint fundamentally limits the effectiveness of perturbation design, we further observe that existing proactive defenses face broader applicability constraints. As shown in Table \ref{tab:SOTA_with_ours}, most methods are tailored to GAN-based attribute editing and seldom address more challenging manipulations such as face-swapping, let alone identity-preserving facial generation. Moreover, many approaches assume full white-box access to model gradients, which is unrealistic because most deepfake systems are closed cloud services. 
Existing black-box defenses still focus exclusively on GANs, offering no protection against the rapidly growing diffusion-based deepfake models. These limitations highlight a clear gap: \textit{a practical defense should function with or without gradient access, generalize to multiple manipulation types, and remain effective on both GAN and diffusion-based generators.} 

To overcome these limitations, we propose AEGIS, the first \textit{diffusion-guided} paradigm in which \textbf{A}dv\textbf{E}rsarial facial images are \textbf{G}enerated for \textbf{I}dentity \textbf{S}hielding, enabling generalizable deepfake defense. Instead of directly optimizing pixel perturbations under a fixed $L_\infty$ bound, AEGIS progressively injects adversarial signals into the denoising trajectory of a DDIM diffusion model. This formulation inherently avoids the peak-clipping constraint imposed by traditional perturbation optimization and enables perturbation strength to self-adapt to semantically sensitive facial regions. Moreover, unlike prior defenses that rely on gradient access or perturbation transferability, AEGIS functions in both white-box and black-box settings by leveraging gradient-based guidance or gradient-free estimation, respectively. The result is a unified and training-free defense mechanism capable of disrupting GAN- and diffusion-based deepfake generation across attribute editing, face swapping, and identity-preserving synthesis.

In summary, this work makes the following key contributions:
\begin{itemize}
    \item \textbf{New Paradigm.}
We propose AEGIS, the first diffusion-guided paradigm that generates adversarial facial images for identity shielding. Unlike prior defenses that rely on model internals or additional training, AEGIS performs perturbation generation without model training and functions in both white-box and black-box settings.

\item \textbf{Novel Perspective.}
Instead of accepting the $L_\infty$-bounded paradigm as an unquestioned premise like all prior works, AEGIS rethinks how adversarial perturbations should be formed. We introduce a new perturbation-injection perspective that injects adversarial signals into the diffusion denoising trajectory, enabling perturbations to self-adapt to semantic importance without explicit clipping.

\item \textbf{Broad Applicability.} AEGIS is applicable across diverse deepfake manipulation types and threat environments. It supports attribute editing, face swapping, and identity-preserving facial generation, and works on both GAN-based and diffusion-based deepfake generators under either white-box or black-box settings.

\item \textbf{Strong Performance.}
Extensive experiments show that AEGIS consistently achieves superior defense effectiveness while keeping perturbations visually imperceptible. Furthermore, the generated adversarial images retain minimal identity-recognizable information, effectively mitigating facial stigmatization and reducing the risk of identity misuse.

\end{itemize}

\begin{table*}
\caption{Comparison of the Proposed Method with SOTA Defenses across Multiple Application Scenarios}
\centering
\label{tab:SOTA_with_ours}
\begin{tabular}{ccccccccc}
\hline
Methods & Venue& \begin{tabular}[c]{@{}c@{}}Attribute\\ editing\end{tabular} & \begin{tabular}[c]{@{}c@{}}Face-\\ swapping\end{tabular} &\begin{tabular}[c]{@{}c@{}}ID-preserving\\ facial generation\end{tabular} & \begin{tabular}[c]{@{}c@{}}White-box\\ Setting\end{tabular} & \begin{tabular}[c]{@{}c@{}}Black-box\\ Setting\end{tabular} &\begin{tabular}[c]{@{}c@{}}GAN-based\\ Manipulation\end{tabular} & \begin{tabular}[c]{@{}c@{}}Diffusion-based\\ Manipulation\end{tabular} \\
\hline
White-Blur\cite{2020_white_blur}& ECCV 2020 & $\checkmark$ & \ding{55} & \ding{55}&$\checkmark$ & \ding{55} &  $\checkmark$& \ding{55} \\
Spread-Spectrum\cite{2020_white_blur} &ECCV 2020 & $\checkmark$ & \ding{55} & \ding{55} &$\checkmark$ & \ding{55} &$\checkmark$  & \ding{55} \\
Anti-Forgery\cite{ijcai2022p107} & IJCAI 2022&$\checkmark$ & \ding{55} &\ding{55} & $\checkmark$ &\ding{55}  &$\checkmark$   & \ding{55} \\
Saliency-Aware\cite{2024_saliency_aware} &TCSS 2024&  $\checkmark$  & \ding{55} &\ding{55} & $\checkmark$   &\ding{55}  &  $\checkmark$  & \ding{55} \\
Union-Aware\cite{2024_union-saliency}&TCE 2024 &$\checkmark$   &\ding{55}   & \ding{55} &$\checkmark$  &  \ding{55} & $\checkmark$  &\ding{55}   \\
DF-RAP\cite{2024_DF_RAP} &TIFS 2024& $\checkmark$  & $\checkmark$  & \ding{55} &$\checkmark$  & \ding{55}  & $\checkmark$& \ding{55} \\
Venom\cite{huang2021initiative} & AAAI 2021& $\checkmark$  & \ding{55}  &\ding{55} & \ding{55}  & $\checkmark$  & $\checkmark$& \ding{55} \\
TCA-GAN\cite{2023_TIFS_Black_box} &TIFS 2023&  $\checkmark$ &  $\checkmark$ &\ding{55} &\ding{55}  &  $\checkmark$ &$\checkmark$  & \ding{55}\\
RUIP\cite{zhang2025robust}&TIFS 2025&  $\checkmark$ &  $\checkmark$ &\ding{55} &\ding{55}  &  $\checkmark$ &$\checkmark$  & \ding{55}\\
AEGIS &---&$\checkmark$  & $\checkmark$ & $\checkmark$ &$\checkmark$ & $\checkmark$ & $\checkmark$ &$\checkmark$ \\
\hline
\end{tabular}
\end{table*}

\section{Related Work}
\subsection{Facial Manipulation Models}
Deepfake facial manipulation models are generative models that modify or synthesize facial images while maintaining photorealism and identity consistency. Based on their manipulation objectives, these models can be categorized into three major groups: facial attribute editing, face swapping, and identity-preserving facial generation. 
Facial attribute editing models modify semantic facial attributes (e.g., age, hair color, expression) of an input face while preserving the identity of a person and overall facial structure. Representative methods include StarGAN~\cite{stargan}, AttGAN \cite{attgan}, HiSD \cite{HiSD}, FPGAN \cite{FPGAN}, and AttentionGAN \cite{tang2021attentiongan}.
Face-swapping models transfer the identity information from a source face onto a target face while maintaining the target’s pose, expression, and background context. Typical examples include SimSwap~\cite{simswap} and Faceshifter\cite{li2020advancing}. 
Identity-preserving facial generation models synthesize new facial images conditioned on identity embeddings extracted from an input face, ensuring the generated images retains identity semantics while allowing free variation in visual appearance, such as pose, illumination, or style. Recent works such as Arc2face\cite{papantoniou2024arc2face} adopt diffusion models to achieve high-fidelity identity-consistent generation. Our proposed AEGIS operates effectively across all three categories, providing a unified proactive defense method.

\subsection{Proactive Defense against Facial Manipulation}
Proactive adversarial defenses against facial manipulation can be categorized into two settings based on the defender’s knowledge of the manipulation model. In the \textit{white-box setting}, the defender has full access to the deepfake generator, including its architecture, parameters, and gradients, enabling direct optimization of adversarial perturbations through back-propagation. In the \textit{black-box setting}, the internal information of the deepfake generator is inaccessible; the defender can only query the model outputs and must generate perturbations without gradient access.

For a long time, research on proactive deepfake defense has primarily focused on improving defense effectiveness under the \textbf{white-box setting}. An early representative work, White-Blur~\cite{2020_white_blur}, introduced the notion of disrupting deepfakes and adapted adversarial attack techniques into facial attribute-editing pipelines. The same work further proposed a Spread-Spectrum perturbation strategy to withstand simple blur-based countermeasures. Yeh et al.~\cite{attack_as_defense} extended this direction by presenting a limit-aware adversarial attack, which exploits the numerical range constraints inherent in image-to-image translation GANs. Anti-Forgery\cite{ijcai2022p107} further pursues perceptual-aware perturbations that remain stealthy while degrading manipulation fidelity. Subsequent works refined where and how to allocate the perturbation budget. Saliency-Aware \cite{2024_saliency_aware} biases gradients toward semantically important facial regions to maximize disruption with limited distortion, while Union-Aware\cite{2024_union-saliency} aggregates multi-scale saliency priors to stabilize performance across diverse GAN manipulators. To address degradation on social platforms, DF-RAP\cite{2024_DF_RAP} explicitly optimizes perturbations that survive online post-processing. In parallel, CMUA \cite{Huang2021CMUAWatermarkAC} proposes a cross-model universal adversarial watermark by jointly optimizing against multiple GAN manipulators, improving model coverage under a single perturbation.

In contrast to white-box settings, research on proactive defense in \textbf{black-box scenarios} remains relatively limited. Huang et al. \cite{huang2021initiative} took an early step toward black-box applicability by proposing initiative defense, in which perturbations are generated using a surrogate manipulation model to degrade real deepfake outputs. Although the method weakens dependence on model internals, it still relies on training a surrogate generator and therefore does not fully disengage from gradient accessibility. Dong et al. \cite{2023_TIFS_Black_box} enhanced black-box defense by generating highly transferable perturbations using a substitute reconstruction model, allowing manipulation disruption without gradient access or model queries. More recently, Zhang et al. \cite{zhang2025robust} focused on generating imperceptible perturbations that not only disable manipulation but also avoid leaving identifiable facial traces in the perturbed image.

Despite these advances, both white-box and black-box approaches share fundamental limitations. Nearly all existing defenses optimize perturbations under a fixed 
$L_\infty$ bound to maintain imperceptibility, which inevitably weakens defense strength. Meanwhile, research in this domain remains fragmented; no prior work is able to operate consistently across both white-box and black-box settings. Although the recent RUIP~\cite{zhang2025robust} avoids hard $L_\infty$ constraints by using a GAN to generate adversarial faces, it requires balancing multiple loss terms to preserve visual quality and defense effectiveness, resulting in expensive training overhead and limited scalability. In contrast, our AEGIS is the first framework that simultaneously removes the $L_\infty$-bounded peak-clipping constraint and eliminates the need to train a generative model, enabling a unified, training-free, and generalizable defense applicable to both white-box and black-box deepfake systems.

\section{Motivation and Threat Model}

\subsection{Motivation}
\label{sec.motivation}

In a proactive defense scenario, the defender deliberately injects a perturbation $\eta$ into the original facial image $x$. The resulting adversarial image $x^{adv}=x+\eta$ is designed to cause the deepfake model $M$ to produce a synthesized facial image $M(x^{adv})$ that appears less realistic or is semantically inconsistent with the original identity. In this scenario, two primary objectives must be jointly addressed: \textit{defense effectiveness} and \textit{perturbation imperceptibility}. The former evaluates how effectively the adversarial perturbation disrupts the deepfake model’s output, while the latter requires that the perturbation remain visually imperceptible to human observers.

\textit{\uline{Breaking the Peak-clipping Constraint}}: To achieve defense effectiveness, regardless of the specific perturbation-generation strategy employed, most existing approaches have a common optimization objective formulated as
\begin{equation}
\max_{\|\eta\|_{\infty} \leq \varepsilon} L(M(x), M(x+\eta)),
\label{eq:objective_of_SOTA}
\end{equation}
where \(L(\cdot)\) denotes a distance metric (e.g., MSE) to quantify the degradation of the manipulated output, and \(\varepsilon\) denotes the upper bound of the perturbation magnitude under the $L_\infty$ norm. 
However, the $L_\infty$-bounded optimization paradigm adopted by most existing methods \cite{2020_white_blur,ijcai2022p107, 2024_saliency_aware,2024_union-saliency,2024_DF_RAP} inherently suffers from a \textit{peak-clipping} problem. The uniform constraint $|\eta_i| \leq \varepsilon$ clips high-gradient regions that require stronger perturbations while still adding unnecessary noise to less sensitive areas, resulting in spatial energy imbalance and inefficient perturbation use. Consequently, the overall defense performance becomes suboptimal, as the regions most capable of disturbing deepfake generation are precisely those where the perturbation amplitude is most restricted. Although relaxing this bound could enhance defense strength, it would inevitably weaken perturbation imperceptibility and risk visible artifacts. The key challenge, therefore, is to balance effective manipulation disruption with the preservation of visual imperceptibility.

\textit{\uline{Enhancing the Versatility and Practicality}}: Beyond effectively addressing the peak-suppression problem, this work is also motivated by the need for greater versatility and practicality in proactive deepfake defense. First, most existing methods are \textit{too narrow in scope}—they are specifically designed for attribute editing, a relatively simple and controllable manipulation, while overlooking more complex and realistic deepfake scenarios such as face swapping and identity-preserving facial generation. Second, current defenses are \textit{too idealized in gradient accessibility}, as they typically assume a white-box setting where model gradients are available, an assumption rarely valid in real-world applications. In practice, deepfake generators often operate as black-box systems, making gradient-based defenses infeasible. Third, existing methods are \textit{too outdated in model coverage}, focusing almost exclusively on GAN-based deepfake models despite the rapid emergence of diffusion-based generation techniques. Therefore, as presented in Table \ref{tab:SOTA_with_ours}, our goal is to develop a unified defense framework that remains effective across both white-box and black-box settings and can robustly defend against diverse deepfake types, including GAN- and diffusion-based facial manipulation.

\subsection{Threat Model}

%\begin{quote}
%``\textit{Know your enemies and know yourself, you will not be %imperiled in a hundred battles.}'' \\
%--- Sun Tzu, \textit{The Art of War}
%\end{quote}

We study a proactive defense scenario where the defender releases a protected facial image to mitigate subsequent deepfake manipulation. Let $x$ denote the original facial image owned by the defender. The defender publishes only a perturbed version $x^{adv} = x + \eta$, while keeping $x$ private.  The adversary (deepfaker) is assumed to have access to $x^{adv}$ but not the clean $x$; otherwise, the adversary can simply manipulate $x$ directly and any input-side proactive defense becomes moot.

\subsubsection{The Goal and Abilities of Adversary}

Given $x^{adv}$, the adversary aims to produce a manipulated output that is visually realistic and satisfies the intended manipulation objective, such as attribute editing, face swapping, or identity-preserving synthesis. To maximize success, the adversary may employ an arbitrary facial manipulation model $M(\cdot)$, including both GAN- and diffusion-based generators, and can switch across models. The adversary may further apply standard input transformations to $x^{adv}$ (e.g., resizing, compression, blurring, or denoising) as countermeasures prior to manipulation. The deepfake is considered successful if the manipulated output $M(x^{adv})$ preserves the desired semantic attributes with high perceptual fidelity. 

\subsubsection{The Goal and Abilities of Defender}

We do not assume that the defender can modify, watermark, or interfere with the deepfake model itself; the only controllable variable is the released input image. 
Hence, the defender's objective is to produce $x^{adv}$ so that the resulting output $M(x^{adv})$ becomes semantically inconsistent,identity-invalid, or otherwise unusable, while keeping $x^{adv}$ visually indistinguishable from $x$.

We consider two defender-knowledge settings. In the \textbf{white-box setting}, the defender has access to the model architecture and parameters of $M(\cdot)$, allowing exact gradient computation with respect to the adversarial objective. In this case, the perturbation direction is derived from the gradient of the adversarial loss, i.e.,
\begin{equation}
\eta \propto \nabla_x L_{adv}(M(x)).
\end{equation}

In the \textbf{black-box setting}, the defender has no access to model parameters, architecture, or gradients. Only the model outputs can be observed. To obtain adversarial guidance, the gradient direction is estimated in a query-efficient manner using finite differences over model outputs:
\begin{equation}
\nabla_x M(x) \approx \frac{M(x + \delta) - M(x - \delta)}{2\delta},
\end{equation}
where $\delta$ is an input perturbation probe. The estimated direction is then injected into the diffusion denoising trajectory to synthesize adversarially protected images. Thus,  the defense remains applicable even when the deepfake generator is deployed as a closed, remote inference service.

Across both settings, the defender aims to synthesize $x^{adv}$ such that deepfake manipulation fails, i.e.,
$M(x^{adv}) \not\approx M(x)$,
while preserving imperceptibility and eliminating recognizable identity cues to reduce the risk of identity misuse.

\section{Methodology}
\subsection{Preliminaries and Overview  of AEGIS}
Motivated by the objectives discussed above, we propose the method \textbf{AEGIS}, a method that generates adversarial face images for generalizable defense against facial manipulation.  To address the peak-cropping constraint while maintaining visual imperceptibility, we employ a diffusion model as the guiding generator. Unlike $L_\infty$-bounded optimization that directly clips excessive perturbations, the diffusion process performs gradual noise refinement through iterative denoising, softly regulating perturbation strength rather than discarding it. This property allows controlled enhancement in semantically important regions while preserving visual fidelity. Moreover, diffusion models naturally provide a flexible generative prior, enabling our framework to generalize across multiple deepfake manipulation types and both white-box and black-box settings without additional adversarial training.

\begin{figure*}
    \centering
    \includegraphics[width=1\linewidth]{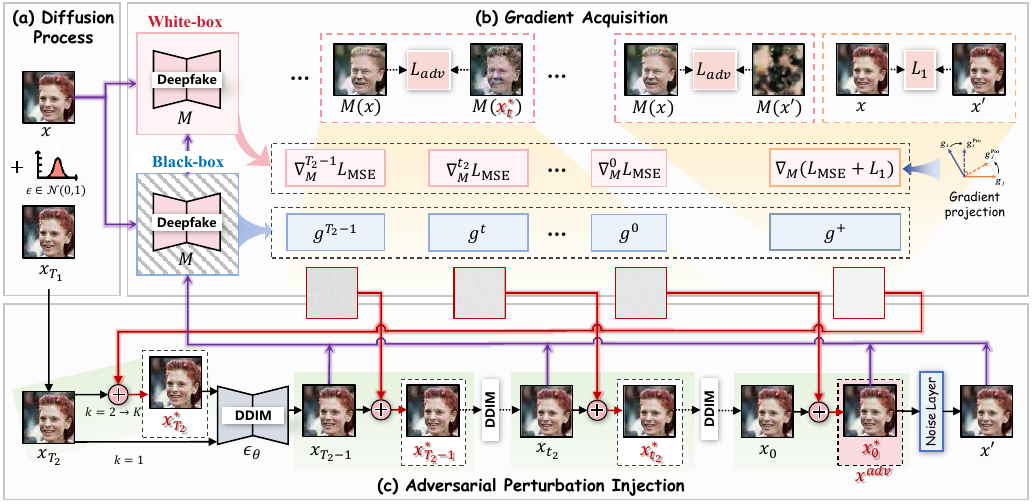}
    \caption{Overview of AEGIS. AEGIS generates adversarial facial images by injecting perturbations into the DDIM denoising trajectory. (a) Diffusion Process: the input face $x$ is forward-diffused to a noisy latent $x_{T_1}$. (b) Gradient Acquisition: obtain gradients from the deepfake model—directly in white-box, or via gradient estimation in black-box—and compute the perturbation from these gradients. (c) Adversarial Perturbation Injection: the computed perturbations are progressively added at selected DDIM denoising steps to guide reconstruction toward the final adversarial output $x^{adv}$. }
    \label{fig:Method_Overview}
\end{figure*}

In our proposed method, DDIM\cite{DDIM} is employed as the guiding generative model. DDIM consists of a forward diffusion process and a reverse denoising process. In the forward process, a clean image $x_0$ is gradually perturbed by adding Gaussian noise $\epsilon \in \mathcal{N}(0,1)$ through a Markov chain parameterized by a noise schedule $\{\beta_{t_1}\}_{t_1=1}^{T_1}$. The latent variable $x_{T_1}$ at diffusion step $t_1$ is defined as:
\begin{equation}
\label{eq_forward}
\setlength{\abovedisplayskip}{0.5mm}
\setlength{\belowdisplayskip}{0.5mm}
    q(x_{t_1}|x_{t_1-1}) = \mathcal N(x_{t_1}; \sqrt{1 - \beta_{t_1}} x_{t_1-1}, \beta_{t_1}\mathbf{I}),
\end{equation}
where $\mathbf{I}$ denotes the identity matrix. Let $\alpha_{t_1}=1-\beta_{t_1} $,$\bar{\alpha_{t_1}}=\prod^{t_1}_{i=1}\alpha_i$; by reparameterization, $x_{t_1}$ can be expressed as $x_{t_1}=\sqrt{\bar{\alpha_{t_1}}}x_0+\sqrt{1-\bar{\alpha_{t_1}}}\epsilon$.

The reverse denoising process aims to reconstruct $x_0$ from the noise latent $x_{T_2}$ by iteratively removing the added noise. The reverse update at step $t_2$ is formulated as:
\begin{equation}
\label{eq_backward}
    x_{t_2-1}=\sqrt{\bar{\alpha}_{t_2-1}}x_0+\sqrt{1-\bar{\alpha}_{t_2-1}}\epsilon_\theta(x_{t_2},t_2),
\end{equation}
where $\epsilon_\theta(x_{t_2},t_2)$ denotes the predicted noise component from the denoising network. In practice, $x_0$ can be estimated directly from $x_{t_2}$ via:
\begin{equation}
    x_0=\frac{x_{t_2}-\sqrt{1-\bar{\alpha}_{t_2}}\epsilon_\theta(x_{t_2},t_2)}{\sqrt{\bar{\alpha}_{t_2}}}.
\end{equation}

Given the DDIM structure, to synthesize adversarial facial images without retraining the generative model, it is preferable to inject perturbations during the reverse denoising process rather than in the forward noising phase. The forward diffusion process is designed to destroy image structure by continuously adding Gaussian noise, driving the data distribution toward an isotropic prior. Adversarial perturbations introduced at this stage are easily neutralized during subsequent denoising and thus fail to meaningfully influence the final synthesis. Conversely, the reverse process reconstructs semantic details step by step, making it an ideal stage for adversarial manipulation. By introducing perturbations into selected denoising steps, we can directly interfere with the model’s semantic reconstruction trajectory, steering the generation toward visually consistent yet semantically disrupted results.

Hence, as illustrated in Fig.~\ref{fig:Method_Overview}, AEGIS injects adversarial perturbation into the DDIM denoising process. The original facial image $x$ is first treated as the input clean image $x_0$ and diffused to a noisy latent $x_{T_1}$. During the reverse denoising phase, adversarial perturbations are progressively injected into selected steps,guiding the reconstruction until the adversarial output $x^{adv}$. By injecting perturbations in latent space rather than directly onto pixels, AEGIS avoids the peak-clipping constraint and allows perturbation magnitude to self-adapt during denoising. In the white-box setting, adversarial perturbations are computed directly from the deepfake model; in the black-box setting, the perturbations are obtained via the gradient-estimation strategy.

\subsection{Adversarial Defense of AEGIS }
As defenders, in practical deepfake defense scenarios, we may possess varying levels of knowledge about the targeted facial manipulation model. In this work, we consider two representative settings based on the background knowledge of the defender. In the white-box scenario, the defender has full access to the deepfake generation model, including its architecture and parameters. In contrast, the black-box scenario assumes no knowledge of the model internals, and only the generated outputs are observable. For the proposed AEGIS framework, the forward diffusion process used to obtain $x_{T_1}$ follows Eq.~(\ref{eq_forward}) identically in both settings. The key distinction lies in the design of the reverse denoising process, which determines how adversarial facial images are generated under each scenario. Accordingly, we develop dedicated defense strategies for both white-box and black-box conditions, while employing DDIM as the unified guiding generative model.

\subsubsection{White-box Scenario} AEGIS for white-box scenario has three key parts. They are adversarial perturbation injection, utilizing noise layer and gradient projection. The whole procedure is presented in Algorithm \ref{algo:AEGIS}. The details of each component are elaborated as follows.

\uline{\textit{Injecting Adversarial Perturbation}}:
To synthesize adversarial facial images using DDIM, perturbations are injected progressively during the denoising process. To maintain the visual imperceptibility, the perturbation is applied step by step from $x_{T_2}$ down to $x_0$. At an arbitrary denoising step $t_2\in [T_2, 1]$, the perturbation result $x_{t_2-1}^*$ is formulated as:
\begin{equation}
    x_{t_2-1}^*= x_{t_2-1}+\alpha \nabla L_{adv},
\end{equation}
where $x_{t_2-1}$ is computed following Eq.~(\ref{eq_backward}), $\alpha$ is a guidance parameter controlling the perturbation strength, and $\nabla L_{adv}$ represents the gradient of the adversarial objective.
The defense objective aims to maximize the discrepancy between the original and manipulated outputs produced by the deepfake model $M$. Accordingly, we adopt the Mean Squared Error (MSE) as the adversarial loss:
\begin{equation}
    L_{adv}=L_{\text{MSE}}=\|M(x) - M(x_{t_2-1})\|_2^2,
\end{equation}
where $x$ is the original facial image and $x_{t_2-1}$ is the partially denoised result at step $t_2$ .
In the white-box scenario, since the gradient of $M$ is accessible, the perturbation updated at step $t_2$ can be written as:
\begin{equation}
    x_{t_2-1}^*= x_{t_2-1}+\alpha \nabla_M L_{\text{MSE}}(M(x), M(x_{t_2-1})).
\end{equation}
This iterative process ensures that the adversarial guidance is introduced coherently throughout the denoising trajectory, progressively steering the generation toward semantically distorted but visually imperceptible adversarial facial images.

\textit{\uline{Utilizing Noise Layer}}: Diffusion-based denoising naturally refines perturbations over successive steps, allowing them to be progressively stabilized by the learned score prior.
As a result, perturbations injected within the DDIM denoising trajectory are not simply amplified or clipped; instead, they are softly regularized toward perceptually coherent directions.
This enables the adversarial perturbation to preserve its disruptive effect even after the resulting image undergoes degradation such as compression or smoothing.

To further enhance robustness against real-world post-processing, we introduce a noise layer during generation. After obtaining $x^*_{t_2-1}$ at step $t_2$, we simulate post-processing by applying a lightweight distortion:
\begin{equation}
    x'=\mathcal{T}(x^*_{t_2-1}),
\end{equation}
where $\mathcal{T}(\cdot)$ corresponds to Gaussian smoothing in our implementation.

\textit{\uline{Gradient Projection}}: The disturbed result $x'$ is then fed into the deepfake model $M$ to compute gradients for adversarial guidance.
Specifically, we obtain the gradient $g_1$ for disrupting the deepfake output and $g_2$ for preserving visual fidelity as:
\begin{align}
g_1 &= \nabla_M L_{\text{MSE}}\big( M(x), M(x') \big), \\
g_2 &= \nabla_M L_{1}\big( x, x' \big),
\end{align}
where $L_1(\cdot,\cdot)$ denotes the pixel-wise $L_1$ distance. 

Although the two gradients pertain to visual perception, they serve opposite purpose: $g_1$ drives the adversarial objective by maximizing the deepfake output discrepancy, whereas $g_2$ preserves visual similarity to maintain imperceptibility. Because these objectives are inherently conflicting, jointly optimizing $g_1$ and $g_2$
may lead to gradient interference, which can weaken the defense effectiveness. To address this issue, we adopt the gradient cross-projection strategy from \cite{GRASP}. 

When the inner product of two gradients is non-positive—indicating conflicting directions—we project one gradient onto the orthogonal complement of the other. The resulting merged gradient is defined as: 
\begin{equation}\label{eq:gradient_projection}
G(g_1, g_2) =
\begin{cases}
    -\lambda_1\, \text{Proj}_{g_2} g_1 + \mu_1\, \text{Proj}_{g_1} g_2 , & \text{if } \langle g_1, g_2 \rangle \leq 0, \\
    -\lambda_2\, g_1 + \mu_2\, g_2, & \text{otherwise},
\end{cases}
\end{equation}   
where the projection onto the normal plane is given by
\begin{equation}
    \text{Proj}_{g_2} g_1 = g_1 - \frac{\langle g_1, g_2 \rangle}{\|g_2\|^2} g_2, \quad 
    \text{Proj}_{g_1} g_2= g_2 - \frac{\langle g_2, g_1 \rangle}{\|g_1\|^2} g_1,
\end{equation}
\(\lambda_1\), \(\lambda_2\), \(\mu_1\) and \(\mu_2\) are hyperparameters controlling the contribution of each gradient.
    
Instead of performing the denoising only once, AEGIS repeats this perturbation-guided denoising procedure for $K$ iterations.
Across iterations, each step refines the adversarial signal while preserving visual fidelity, and the perturbation progressively accumulates along the denoising trajectory until the final adversarial face image $x^{adv}$ is obtained.

\begin{algorithm}[h]
\caption{White-box Scenario of AEGIS}
\label{algo:AEGIS}

\KwIn{Original face image $x$; Deepfake generator $M$; 
Guidance weights $\alpha, \lambda_1, \mu_1, \lambda_2, \mu_2$; 
DDIM diffusion step $T_1$ and denoising step $T_2$; Number of iterations $K$.
}
\KwOut{Adversarial face image $x^{adv}$}

\tcp{Diffusion Procedure}
$x_{T_1} \leftarrow \text{DDIMInversion}(x, T_1)$\;
$x_{T_2} \leftarrow x_{T_1}$\;

$x^{tmp} \leftarrow x_{T_2}$\;
\For{$k \leftarrow 1$ \KwTo $K$}{
  % $x_{T_2} \leftarrow x^{tmp}$\;
  $x^{adv} \leftarrow x^{tmp}$\;
  \tcp{Injecting Adversarial Perturbation}
  \For{$t_2 \leftarrow T_2$ \KwTo $1$}{
    $x_{t_2-1} \leftarrow \text{DDIMSample}(x^{adv})$\;
    $g \leftarrow \nabla_M L_{\mathrm{MSE}}\!\big(M(x),\,M(x_{t_2-1})\big)$\;
    $x_{t_2-1}^{*} \leftarrow x_{t_2-1} + \alpha g$\;
    $x^{adv} \leftarrow x_{t_2-1}^*$\;
  }
  % $x^{adv} \leftarrow x_1^*$\;
  \tcp{Adding Noise Layer}
  $x' \leftarrow \text{NoiseLayer}(x^{adv})$\;
  \tcp{Gradient Projection}
  $g_1 \leftarrow \nabla_M L_{\mathrm{MSE}}\!\big(M(x),\,M(x')\big)$\;
  $g_2 \leftarrow \nabla_M L_{1}(x, x')$\;

  \If{$\langle g_1,g_2\rangle < 0$}{
    $x^{tmp}  \leftarrow x_{T_2} - \lambda_1\mathrm{Proj}_{g_2}(g_1) + \mu_1\mathrm{Proj}_{g_1}(g_2)$\;
  }
  \Else{
    $x^{tmp}  \leftarrow x_{T_2} - \lambda_2 g_1 + \mu_2 g_2$\;
  }
  % $x^{tmp} \leftarrow x_0^*$\;
} 
 \textbf{return} $x^{adv}$
\end{algorithm}

\subsubsection{Black-box Scenario}
For the black-box setting of AEGIS, the overall pipeline remains unchanged, where perturbations are still progressively injected during the DDIM denoising process.
The only substantive difference is that the deepfake model $M$ does not expose its parameters or gradients, making direct computation of $\nabla L_{adv}$ infeasible. 

\textit{\uline{Gradient Estimation}}: To overcome this, we adopt the Natural Evolution Strategies (NES)-based gradient estimation method from \cite{ilyas2018black}. NES allows estimating gradients using only model's outputs by evaluating the loss under sampled perturbations. The core idea is to treat gradient estimation as derivative-free optimization, where random Gaussian directions are sampled around the current point, and the gradient is approximated by observing how the loss changes along those directions.
Formally, for an adversarial objective $L_{adv}$, the NES gradient estimator is:
\begin{equation}
\nabla_x L_{adv} \approx \frac{1}{n\sigma}
\sum_{i=1}^{n} \left[ L_{adv}(x + \sigma u_i) - L_{adv}(x - \sigma u_i) \right] u_i,    
\end{equation}
where $u_i \sim \mathcal{N}(0, I)$ represents Gaussian sampling directions and $\sigma$ determines the variance, controlling the magnitude of the perturbation applied along each direction.

\uline{\textit{Injecting Adversarial Perturbation}}: As shown in Algorithm~\ref{alg:AEGIS_blackbox_denoise}, we replaced the gradient calculations with NES estimation, and once the pseudo-gradient is estimated, perturbation injection during DDIM denoising follows the same mechanism as in the white-box case:
\begin{equation}
    x_{t_2-1}^* = x_{t_2-1} + \alpha \hat{\nabla} L_{adv},
\end{equation}
where $ \hat{\nabla} L_{adv}$ is the NES-estimated gradient.
This allows AEGIS to operate entirely without gradient access, ensuring applicability to real-world deepfake systems where the manipulation model is deployed as a closed black-box service. 

\begin{algorithm}[]
\caption{Black-box Scenario of AEGIS}
\label{alg:AEGIS_blackbox_denoise}
\DontPrintSemicolon
\SetKwInput{KwIn}{Input}
\SetKwInput{KwOut}{Output}
\KwIn{
Original image $x$; Black-box deepfake model $M(\cdot)$ (query access only);
Guidance weights $\alpha$;
DDIM diffusion step $T_1$ and denoising step $T_2$; Number of iterations $K$; NES samples $n$, NES scale $\sigma$.
}
\KwOut{Adversarial facial image $x^{adv}$.}

\tcp{Diffusion Procedure}
$x_{T_1} \leftarrow \text{DDIMInversion}(x, T_1)$\;
$x_{T_2} \leftarrow x_{T_1}$\;
$x^{tmp} \leftarrow x_{T_2}$\;

\For{$k \leftarrow 1$ \KwTo $K$}{
  $x^{adv} \leftarrow x^{tmp}$\;
  \tcp{NES zeroth-order Gradient of the Disruption Objective}
  \For{$t_2 \leftarrow T_2$ \KwTo $1$}{
$x_{t_2-1} \leftarrow \text{DDIMSample}(x^{adv})$\;

    $\hat{g} \leftarrow \mathbf{0}$\;
    \For{$i \leftarrow 1$ \KwTo $n$}{
      sample $u_i \sim \mathcal{N}(0, I)$\;
      $q^+ \leftarrow M(x' + \sigma u_i)$;\quad $q^- \leftarrow M(x' - \sigma u_i)$\;
      $\Delta_i \leftarrow L_{adv}(q^+) - L_{adv}(q^-)$\;
      $\hat{g} \leftarrow \hat{g} + \Delta_i\, u_i$\;
    }
   
%    $g_1 \leftarrow \dfrac{1}{n\sigma}\,\hat{g}$\;
%    $g_2 \leftarrow \nabla_{x'} L_{1}(x,\, x')$\;

%    $g \leftarrow g_1 + \gamma\, g_2$\;

%    \tcp{Guided update of the intermediate sample}
    $x_{t_2-1}^* \leftarrow x_{t_2-1} + \alpha g$\;
    $x^{adv} \leftarrow x_{t_2-1}^*$\;
    $x^{tmp} \leftarrow \text{NoiseLayer}(x^{adv})$
  }
}
\Return $x^{adv} $\;
\end{algorithm}
\vspace{-0.7cm}

\section{Evaluation}
\subsection{Experimental Setup}

\textit{1) Datasets:} Three widely used facial image datasets CelebA\cite{celeba}, FFHQ\cite{FFHQ} and LFW\cite{LFW} are utilized for model training and adversarial facial image generation. CelebA contains more than 200,000 celebrity images annotated with 40 facial attributes, providing rich supervision for attribute editing tasks. FFHQ consists of 70,000 high-resolution images with high visual quality and diversity. LFW includes 13,233 real-world facial images with natural variations in pose, lighting, and expression, making it suitable for evaluating generalization in unconstrained environments. During defense evaluation, 100 images are randomly selected from each dataset individually to serve as its testing set.

\textit{2) Facial Deepfake Models:} The proposed method is evaluated under both white-box and black-box settings. For the \textbf{white-box scenario}, we use three facial attribute editing models—StarGAN \cite{stargan}, AttGAN \cite{attgan}, and HiSD \cite{HiSD}—and one face-swapping model, SimSwap \cite{simswap}. Thirteen attributes are manipulated during editing, covering variations in hair style and color, eyeglasses, gender, age, facial hair, and other features strongly associated with human facial characteristics. For the \textbf{black-box scenario}, we evaluate against StarGAN \cite{stargan}, FPGAN \cite{FPGAN}, AttentionGAN \cite{tang2021attentiongan}, and Simswap \cite{simswap}.
We additionally include Arc2Face \cite{papantoniou2024arc2face}, a diffusion-based identity-preserving face generation model that represents a more realistic and challenging application case.
All these facial deepfake models follow the pretrained configurations and training protocols specified in their original papers.

\textit{3) Comparison Methods:} To demonstrate the effectiveness of AEGIS in proactive defense against facial deepfakes, we compare it with six representative white-box adversarial-perturbation defenses: White-Blur (WB)\cite{2020_white_blur}, S-Spectrum (SS)\cite{2020_white_blur}, Anti-Forgery (AF)\cite{ijcai2022p107}, Saliency-Aware (SA)\cite{2024_saliency_aware}, Union-Aware (UA)\cite{2024_union-saliency}, and DF-RAP (DR)\cite{2024_DF_RAP}. Among them, DF-RAP is the only method that explicitly claims support for proactive defense against face swapping. The remaining five methods were originally designed only for attribute editing, and have no implementation or discussion on defending face-swapping. For a fair comparison, we extend these methods to the face-swapping setting using their loss formulations.
For the black-box setting, we include three representative black-box approaches: Venom \cite{huang2021initiative}, TCA-GAN (TCA) \cite{2023_TIFS_Black_box}, and RUIP \cite{zhang2025robust}, which represent the training-based perturbation-transfer strategy. It is important to note that all existing defenses—both white-box and black-box—are designed exclusively for GAN-based manipulation. None of them address the rapidly emerging diffusion-based identity-preserving facial generation, and thus no prior method can be directly compared in this setting. 

\textit{4) Evaluation Metrics:} To evaluate defense effectiveness, Defense Success Rate (DSR) serves as the primary metric. Its computation varies across application scenarios. For facial attribute editing, following the criterion established in \cite{2020_white_blur}, a defense is considered successful if the \(L_2\) distance between the original and adversarial outputs of the deepfake model \(M\) exceeds 0.05. 
Given that face-swapping introduces sparse and localized changes, pixel-wise norms such as $L_1$ can be reported as auxiliary distortion measures but are not reliable indicators of identity preservation.
Accordingly, in the face-swapping scenario, we adopt the ID sim. metric proposed in Arcface\cite{arcface} to evaluate the DSR. This metric measures the cosine similarity between identity embeddings extracted by the ArcFace model. Following \cite{2024_DF_RAP}, a defense is counted as successful when the ID sim. between the original image and the generated output is below 0.4. In summary, the DSR is:

\begin{equation}
\text{DSR} =
\begin{cases}
\begin{array}{@{}l@{\;}l@{}}
\frac{1}{N} \displaystyle \sum_{i=1}^{N}
\mathbb{I}\!\left( \lVert M(x_i) - M(x_i^{adv}) \rVert_2 > 0.05 \right),
\\ ~~~~~~~~~~~~~~~~~~~~~~~~~~~\text{attribute editing,} \\
\frac{1}{N} \displaystyle \sum_{i=1}^{N}
\mathbb{I}\!\left( \text{ID\_sim}(x_i, M(x_i^{adv})) < 0.4 \right),
\\~~~~~~~~~~~~~~~~~~~~~~~~~~~~~\text{face-swapping},
\end{array}
\end{cases}
\end{equation}
where \(x_i\) and \(x_i^{adv}\) denote the original and adversarial inputs, \(N\) is the number of adversarial images, and \(\mathbb{I}(\cdot)\) is the indicator function. For completeness, $L_1$ is reported in face-swapping experiments as an auxiliary pixel-level metric, but it is not used to compute DSR. Moreover, the Structural Similarity Index Measure (SSIM) is employed to assess the content consistency between the outputs generated from the original and adversarial inputs.

\subsection{White-box Facial Deepfake Scenarios}

In this section, we first evaluate the proposed method under the white-box setting, where the defense algorithm has full access to the internal structure and gradients of the forgery models. This setting allows us to directly optimize adversarial perturbations with respect to the model’s parameters, thereby serving as an upper bound for defense effectiveness. We conduct extensive experiments on both attribute editing and face-swapping models to assess the stability, robustness, and fidelity of the generated adversarial faces. The results demonstrate how well each defense method can suppress forgery generation when complete model information is available.

\begin{table*}[t]
\centering
\caption{Comparison with SOTA Methods on Defense Effectiveness in White-box Settings.}
\label{tab:white-box_DSR}
\begin{tabular}{ccccc|ccc|ccc|ccc}
\toprule
\multirow{2}{*}{Datasets} & \multirow{2}{*}{Methods} &  \multicolumn{3}{c|}{StarGAN~\cite{stargan}} & \multicolumn{3}{c|}{AttGAN~\cite{attgan}} & \multicolumn{3}{c|}{HiSD~\cite{HiSD}} & \multicolumn{3}{c}{SimSwap~\cite{simswap}} \\ \cmidrule(l){3-14} 
 &  & $L_2$$\uparrow$ & SSIM$\downarrow$& DSR$\uparrow$ & $L_2$$\uparrow$ & SSIM$\downarrow$& DSR$\uparrow$ & $L_2$$\uparrow$ & SSIM$\downarrow$ & DSR$\uparrow$ & $L_1$$\uparrow$ & ID sim$\downarrow$ & DSR$\uparrow$ \\ \midrule
\multirow{7}{*}{\begin{tabular}[c]{@{}c@{}}CelebA\\ {\cite{celeba}}\end{tabular}} & WB\cite{2020_white_blur}  & \uline{0.5809} & \uline{0.4582} & \textbf{100.0\%} & \textbf{0.1927} & \textbf{0.7325} & \textbf{89.2\%} & \uline{0.0629} & \uline{0.7673} & 70.0\% & \uline{0.0946} &\uline{0.2650} & \uline{91.5\%} \\
 & SS\cite{2020_white_blur} & 0.3854 & 0.5327 & \textbf{100.0\%} & 0.0928 & 0.8281 & 56.1\% & 0.0476 & 0.8432 & 38.0\% & 0.0842 & 0.2975 & 82.5\% \\
 & AF\cite{ijcai2022p107} & 0.5285 & 0.5260 & \textbf{100.0\%} & 0.0301 & 0.9282 & 17.2\% & 0.0527 &\textbf{0.7559} & \uline{71.0\%} & 0.0581 & 0.4731 & 24.5\% \\
 & SA\cite{2024_saliency_aware} & 0.2833 & 0.6200 & \textbf{100.0\%} & 0.0942 & 0.8338 & 66.4\% & 0.0484 & 0.8541 & 53.0\% & 0.0829 & 0.2757 & 88.0\% \\
 & UA\cite{2024_union-saliency} & 0.2844 & 0.6173 & \uline{99.6\%} & 0.0628 & 0.8901 & 20.8\% & 0.0309 & 0.8737 & 24.0\% & 0.0680 & 0.4066 & 54.0\% \\
 & DR\cite{2024_DF_RAP} & 0.2785 & 0.6031 & \textbf{100.0\%} & 0.1434 & \uline{0.7781} & 71.0\% & 0.0518 & 0.7910 & 50.0\% & 0.0929 & 0.2737 & 83.5\% \\
 &\cellcolor[HTML]{E5E5E5}AEGIS&\cellcolor[HTML]{E5E5E5}\textbf{0.6106} &\cellcolor[HTML]{E5E5E5}\textbf{0.4152} & \cellcolor[HTML]{E5E5E5}\textbf{100.0\%} & \cellcolor[HTML]{E5E5E5}\uline{0.1574} & \cellcolor[HTML]{E5E5E5}0.8375 & \cellcolor[HTML]{E5E5E5}\uline{77.4\%} & \cellcolor[HTML]{E5E5E5}\textbf{0.0930} &\cellcolor[HTML]{E5E5E5}0.8599 & \cellcolor[HTML]{E5E5E5}\textbf{94.5\%} & \cellcolor[HTML]{E5E5E5}\textbf{0.0976} &\cellcolor[HTML]{E5E5E5}\textbf{0.1128} & \cellcolor[HTML]{E5E5E5}\textbf{100.0\%} \\ 
\midrule
\multirow{7}{*}{\begin{tabular}[c]{@{}c@{}}FFHQ\\ {\cite{FFHQ}}\end{tabular}} & WB\cite{2020_white_blur} &\uline{0.5455}&\uline{0.4643} & \textbf{100.0\%} & \textbf{0.2745} & \textbf{0.6664} & \textbf{98.7\%} & \textbf{0.0920} & \uline{0.7898} & \uline{88.5\%} & \uline{0.0890} & 0.3036 & \uline{84.0\%} \\
 & SS\cite{2020_white_blur} & 0.2501 & 0.5965 & \textbf{100.0\%} & 0.1372 & 0.7908 & 83.0\% & 0.0646 & 0.8535 & 63.0\% & 0.0813 & 0.3498 & 72.5\% \\
 & AF\cite{ijcai2022p107} & 0.4301 & 0.5686 & \textbf{100.0\%} & 0.1035 & 0.8501 & 54.6\% & 0.0755 & \textbf{0.7774} & 83.5\%& 0.0514 & 0.5216 & 17.0\% \\
 & SA\cite{2024_saliency_aware} & 0.2577 & 0.6400 & \uline{99.6\%} & 0.1669 & 0.7749 & 89.9\% & 0.0795 & 0.8317 & 84.0\% & 0.0861 & 0.3042 & 79.5\% \\
 & UA\cite{2024_union-saliency} & 0.4255 & 0.5540 & \textbf{100.0\%} & 0.1302 & 0.8042 & 69.4\% & 0.0457 & 0.8796 & 44.5\% & 0.0652 & 0.4334 & 45.5\% \\
 & DR\cite{2024_DF_RAP} & 0.2566 & 0.6118 & \uline{99.6\%} & 0.1726 & 0.7704& 91.5\% & 0.0806 & 0.8047 & 82.0\% & 0.0883 & \uline{0.2907} & 77.5\% \\
 &\cellcolor[HTML]{E5E5E5}AEGIS&\cellcolor[HTML]{E5E5E5}\textbf{0.8127} &\cellcolor[HTML]{E5E5E5}\textbf{0.3294} & \cellcolor[HTML]{E5E5E5}\textbf{100.0\%} & \cellcolor[HTML]{E5E5E5}\uline{0.2744} & \cellcolor[HTML]{E5E5E5}\uline{0.7485} & \cellcolor[HTML]{E5E5E5}\uline{93.9\%} & \cellcolor[HTML]{E5E5E5}\uline{0.0829} &\cellcolor[HTML]{E5E5E5}0.8467 & \cellcolor[HTML]{E5E5E5}\textbf{94.0\%} & \cellcolor[HTML]{E5E5E5}\textbf{0.0989} &\cellcolor[HTML]{E5E5E5}\textbf{0.1233} & \cellcolor[HTML]{E5E5E5}\textbf{99.0\%} \\ 
 \midrule
 \multirow{7}{*}{\begin{tabular}[c]{@{}c@{}}LFW\\ {\cite{LFW}}\end{tabular}} & WB\cite{2020_white_blur} &\uline{0.4893}&\uline{0.4691} & \textbf{100.0\%} & \textbf{0.2759} & \textbf{0.6652} & \textbf{99.8\%} & \textbf{0.0889} &\uline{0.7720} & \uline{93.0\%} & \uline{0.1090} &\uline{0.1572} & \uline{98.0\%} \\
 & SS\cite{2020_white_blur} & 0.3300 & 0.5407 & \textbf{100.0\%} & 0.1392 & 0.7792 & 84.1\% & 0.0718 & 0.8350 & 77.0\% & 0.1007 & 0.1963 & 93.5\% \\
 & AF\cite{ijcai2022p107} & 0.4308 & 0.5395 & \textbf{100.0\%} & 0.1087 & 0.8502 & 55.6\% & 0.0735 & \textbf{0.7689} & 84.0\% & 0.0828 & 0.2685 & 87.5\% \\
 & SA\cite{2024_saliency_aware} & 0.2466 & 0.6248 & \textbf{100.0\%} & 0.1299 & 0.8183 & 85.5\% & 0.0687 & 0.8474 & 85.0\% & 0.0995 & 0.2006 & 95.5\% \\
 & UA\cite{2024_union-saliency} & 0.2720 & 0.6059 & \textbf{100.0\%} & 0.1084 & 0.8227 & 64.5\% & 0.0545 & 0.8508 & 53.5\% & 0.0860 & 0.2742 & 69.5\% \\
 & DR\cite{2024_DF_RAP} & 0.2680 & 0.5744 & \uline{99.2\%} & \uline{0.2340} & \uline{0.6841}& \uline{94.9\%} &\uline {0.0789} & 0.7899 & 87.0\% & \textbf{0.1078} & 0.1842 & 96.5\% \\
 &\cellcolor[HTML]{E5E5E5}AEGIS&\cellcolor[HTML]{E5E5E5}\textbf{0.7212} &\cellcolor[HTML]{E5E5E5}\textbf{0.3510} & \cellcolor[HTML]{E5E5E5}\textbf{100.0\%} & \cellcolor[HTML]{E5E5E5}0.2273 & \cellcolor[HTML]{E5E5E5}0.7361 & \cellcolor[HTML]{E5E5E5}90.9\% & \cellcolor[HTML]{E5E5E5}\textbf{0.0899} &\cellcolor[HTML]{E5E5E5}0.8170 & \cellcolor[HTML]{E5E5E5}\textbf{98.0\%} & \cellcolor[HTML]{E5E5E5}0.1027 &\cellcolor[HTML]{E5E5E5}\textbf{0.1050} & \cellcolor[HTML]{E5E5E5}\textbf{100.0\%} \\ 
\bottomrule
\end{tabular}
\end{table*}

\subsubsection{Defense Effectiveness}
We conduct an experiment to evaluate the overall defense effectiveness and perceptual quality of AEGIS against various white-box facial deepfake models. Table~\ref{tab:white-box_DSR} presents a comprehensive comparison of AEGIS with representative state-of-the-art defense methods, including WB, SS, AF, SA, UA and DR, under white-box conditions across four deepfake models: StarGAN, AttGAN, HiSD, and SimSwap. The evaluation covers both facial attribute editing and face-swapping tasks on three benchmark datasets CelebA, FFHQ, and LFW.

From Table~\ref{tab:white-box_DSR}, we can see that the proposed method AEGIS consistently achieves top-tier defense performance, ranking first or second across almost all the deepfake models. Specifically, we summarized four phenomena: 
\ding{182}AEGIS achieves a 100\% DSR when applied to StarGAN on the CelebA, FFHQ, and LFW datasets, as well as on SimSwap for both CelebA and LFW.
\ding{183}AEGIS achieves significantly higher DSR for HiSD, surpassing the second-best method by 12\%–17\%; this demonstrates its superior defense capability compared to other methods in preventing identity manipulation.
\ding{184}Although AEGIS does not achieve the highest scores on AttGAN, it consistently ranks among the top-performing methods. This is because AttGAN relies heavily on pixel-level integrity for localized attribute editing. Pixel-space defenses directly disrupt the dependency, whereas AEGIS injects perturbations at the semantic level through the denoising process, causing less pixel disruption and thus reduced effectiveness on this model.
\ding{185}In terms of perceptual and distortion metrics, the proposed method yields relatively larger $L_2$ and $L_1$ distances while maintaining moderate SSIM and low ID similarity (ID sim) values; this indicates that the generated perturbations effectively disrupt the identity and attribute consistency of forged images without introducing excessive visual degradation.

\begin{figure*}[t]
    \includegraphics[scale=0.25]{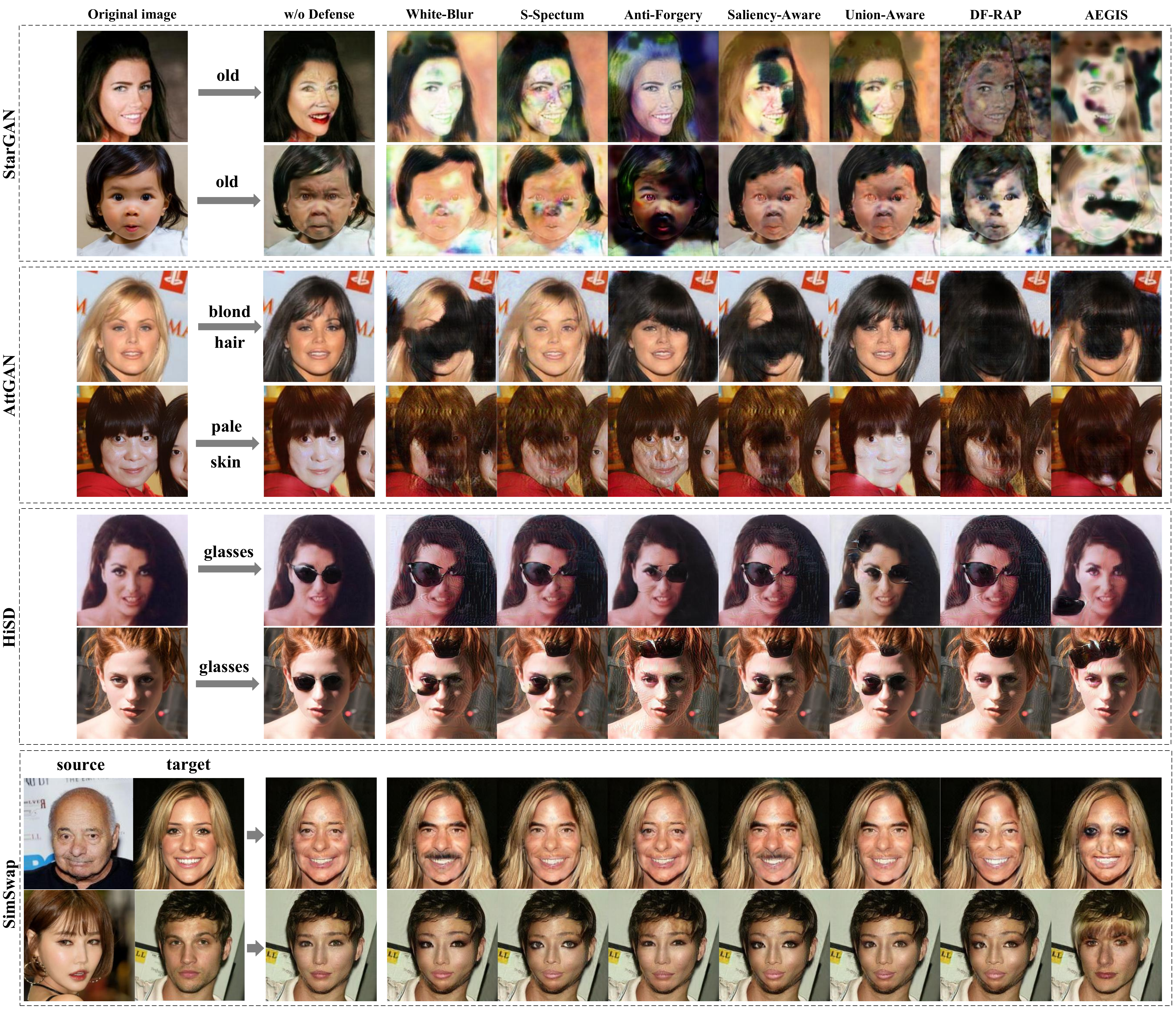}
    \caption{Visualization of Defense effectiveness for SOTA methods and the proposed AEGIS under the white-box scenario.}
    \label{fig:whitebox_dsr_visualization}
\end{figure*}

Figure~\ref{fig:whitebox_dsr_visualization} illustrates the visual defense results for StarGAN, AttGAN, HiSD, and SimSwap, displayed from top to bottom. From the figure we have four key observations: 
\ding{182}For \textbf{StarGAN}, faces protected by AEGIS lose significant structural details during forgery, resulting in visibly distorted outputs that disrupt attribute editing. Specifically, compared to the WB, AF, SA, and UA methods, our approach effectively conceals the face identity to prevent identity misuse in generated images. 
\ding{183}For \textbf{AttGAN}, our method effectively blocks attribute transformation by completely concealing the face identity, making it impossible for the generator to perform meaningful edits. In contrast, SS, AF, and UA either fail to obscure the face or cannot stop the attribute transition, resulting in visible attribute modifications.
\ding{184}In \textbf{HiSD}, while style transformations like hair or skin color are easier to block, adding accessories (e.g., glasses) is more challenging because it introduces a new object rather than altering the existing style. Nevertheless, AEGIS successfully prevents the addition of such objects, whereas competing approaches (WB, SS, AF, SA, UA, and DF-RAP) still generate glasses partially or entirely.
\ding{185}For \textbf{SimSwap}, adversarially protected faces generated by AEGIS show minimal resemblance to the source identity after swapping, while other approaches such as AF still retain recognizable identity cues, demonstrating that our method achieves more thorough and reliable identity concealment.

Overall, these results demonstrate that AEGIS maintains a strong and effective defense, achieving robust and reliable performance across different models and datasets.

\subsubsection{Imperceptibility}
\begin{figure}[]
    \centering
    \includegraphics[width=1\linewidth]{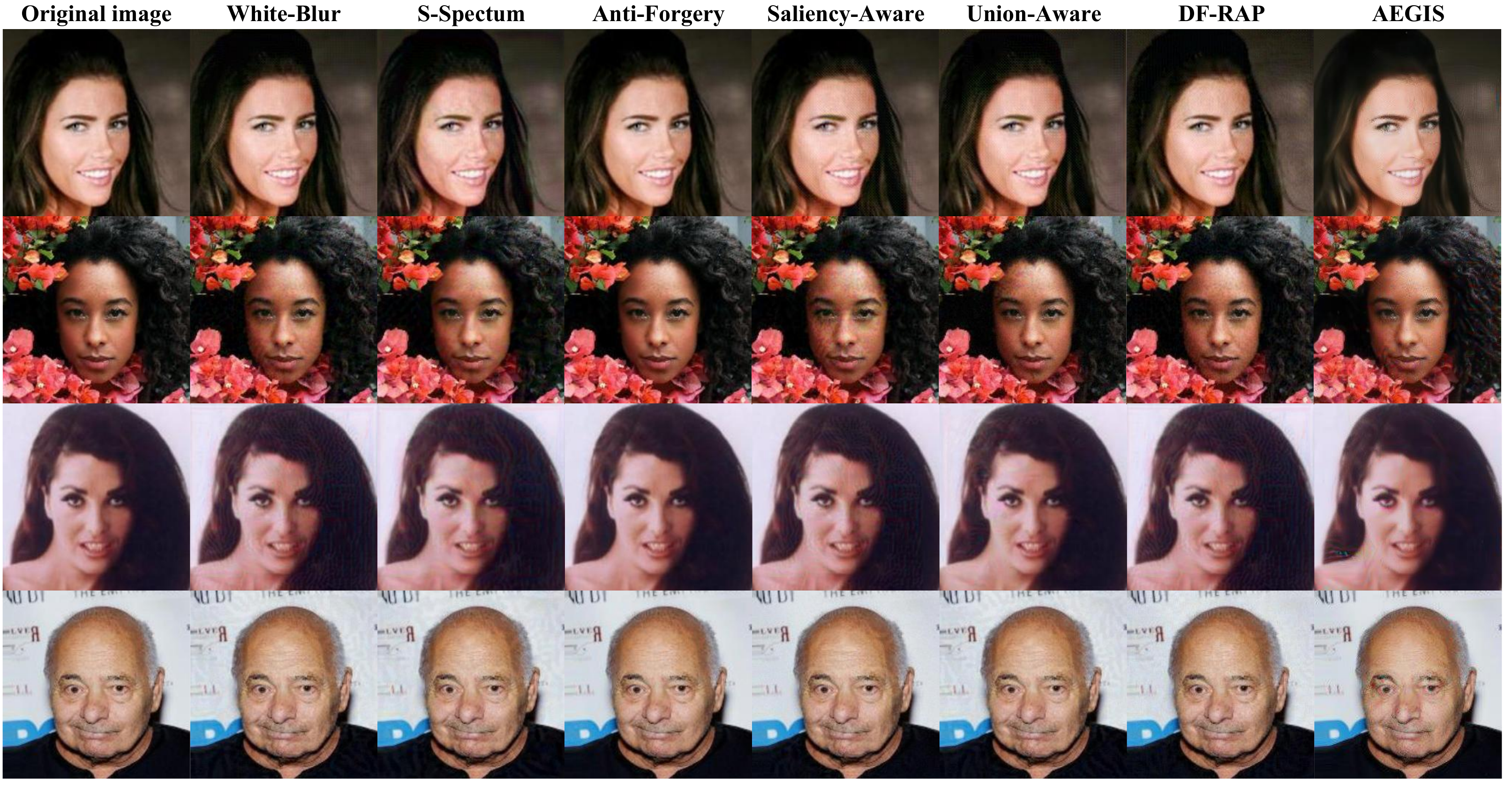}
    \caption{Visualization of perturbation imperceptibility for SOTA methods and AEGIS under the white-box scenario.}
    \label{fig:whitebox_fidelity}
\end{figure}

To evaluate the visual imperceptibility of the adversarial images, that is, the imperceptibility of the adversarial perturbation, a quantitative comparison and visualization of the visual quality of adversarial face images versus the original images are given in Table~\ref{tab:white-box_fidelity} and Figure~\ref{fig:whitebox_fidelity}, respectively. Precisely speaking, we test PSNR, SSIM, and LPIPS \cite{LPIPS} of the adversarial images by comparing them to the original images. LPIPS computes perceptual similarity by measuring feature distances between images using pre-trained deep network.

From Table~\ref{tab:white-box_fidelity}, several key observations can be made. 
First, AEGIS maintains competitive image quality, with PSNR, SSIM, and LPIPS values comparable to or only slightly lower than the best-performing SOTA methods, while consistently attaining the highest DSR across almost all models and datasets. Although AF, SA and UA exhibit the highest perceptual quality in cases of AttGAN, HiSD and SimSwap on CelebA, its defense capability is significantly limited, as reflected by extremely low DSR values (17.0\%$\sim$69.5\%), where high PSNR does not translate into effective defense. 
Therefore, AEGIS demonstrates a more balanced overall performance, achieving competitive visual quality together with the strongest defense results across different models and datasets. 
Furthermore, Figure~\ref{fig:whitebox_fidelity} presents the visual comparison of adversarial samples generated by AEGIS and other defense methods. 
It can be observed that the adversarial faces produced by AEGIS are visually almost indistinguishable from the original images, confirming its strong imperceptibility. 
Finally, the results confirm that simple pixel-level perturbations are insufficient for reliable defense, and the diffusion-guided adversarial mechanism of AEGIS provides a more stable and effective trade-off between invisibility and defense performance.

\begin{table*}[]
\caption{Fidelity Comparison of adversarial images with SOTA methods in white-box settings.}
\label{tab:white-box_fidelity}
\setlength{\tabcolsep}{1.5pt}
\begin{tabular}{@{}c|c|ccc|c||ccc|c||ccc|c||ccc|c@{}}
\toprule
 &  & \multicolumn{4}{c||}{StarGAN\cite{stargan}} & \multicolumn{4}{c||}{AttGAN\cite{attgan}} & \multicolumn{4}{c||}{HiSD\cite{HiSD}} & \multicolumn{4}{c}{SimSwap\cite{simswap}} \\ \cmidrule(l){3-18} 
\multirow{-2}{*}{Datasets} & \multirow{-2}{*}{Methods} & PSNR$\uparrow$ & SSIM$\uparrow$ &LPIPS$\downarrow$ & DSR$\uparrow$ & PSNR$\uparrow$ & SSIM$\uparrow$ & LPIPS$\downarrow$ & DSR$\uparrow$ & PSNR$\uparrow$ & SSIM$\uparrow$ & LPIPS$\downarrow$ & DSR$\uparrow$ & PSNR$\uparrow$ &SSIM$\uparrow$ &LPIPS$\downarrow$ & DSR$\uparrow$ \\ \midrule
 & WB\cite{2020_white_blur} & 34.163 & 0.938 & 0.031 & \textbf{100.0\%} & 33.979 & 0.939 & 0.150 & \textbf{89.2\%} & 33.741 & 0.937 & 0.150 & 70.0\% & 35.040 & 0.955 & 0.072 & \uline{91.5\%} \\
 & SS\cite{2020_white_blur} & 34.298 & \uline{0.956} &0.058 & \textbf{100.0\%} & 32.988 & 0.939 & 0.137 & 56.1\% & 32.916 & 0.952 & 0.109 & 38.0\% & 34.981 & 0.958 & 0.062 & 82.5\% \\
 & AF\cite{ijcai2022p107} & \textbf{53.935} & \textbf{0.999} & \textbf{0.000} & \textbf{100.0\%}& \textbf{43.842} & \textbf{0.992} & \textbf{0.009} & \textcolor{red}{\textbf{17.2\%}} & 31.054 & 0.916 &0.132 & \uline{71.0\%} & \textbf{39.379} & \textbf{0.981} & \textbf{0.0182} & \textcolor{red}{\textbf{24.5\%}} \\
 & SA\cite{2024_saliency_aware} & 34.694 & 0.955 & \uline{0.026} & \textbf{100.0\%} & \uline{34.702} & \uline{0.957} & \uline{0.099} & 66.4\% & 35.775 & 0.971 &0.062 & 53.0\% & \uline{35.783} & 0.961 & \uline{0.059} & 88.0\% \\
 & UA\cite{2024_union-saliency} & 34.612 & 0.951 & 0.028 & \uline{99.6\%} & 30.279 & 0.866 &0.221 & 20.8\% & \textbf{35.803} & \textbf{0.962} & \textbf{0.090} & \textcolor{red}{\textbf{24.0\%}} & 35.496 & 0.958 & 0.073 & 54.0\% \\
 & DR\cite{2024_DF_RAP} & \uline{35.173} & 0.950 & 0.097 & \textbf{100.0\%} & 34.654 & 0.945 & 0.173 &71.0\% & 33.316 & 0.940 & 0.175 & 50.0\% & 35.529 & 0.954 & 0.107 & 83.5\% \\
\multirow{-7}{*}{\begin{tabular}[c]{@{}c@{}}CelebA\\ \cite{celeba}\end{tabular}} & \cellcolor[HTML]{E5E5E5}AEGIS & \cellcolor[HTML]{E5E5E5}32.025 & \cellcolor[HTML]{E5E5E5}0.928 & \cellcolor[HTML]{E5E5E5}0.100 & \cellcolor[HTML]{E5E5E5}\textbf{100.0\%} & \cellcolor[HTML]{E5E5E5}31.142 & \cellcolor[HTML]{E5E5E5}0.932 & \cellcolor[HTML]{E5E5E5}0.158 & \cellcolor[HTML]{E5E5E5}{\uline{77.4\%}} & \cellcolor[HTML]{E5E5E5}{\uline{34.417}} & \cellcolor[HTML]{E5E5E5}{\uline{0.971}} & \cellcolor[HTML]{E5E5E5}{\uline{0.095}} & \cellcolor[HTML]{E5E5E5}{\textbf{94.5\%}} & \cellcolor[HTML]{E5E5E5}34.087 & \cellcolor[HTML]{E5E5E5}{\uline{0.986}} &\cellcolor[HTML]{E5E5E5}0.070 & \cellcolor[HTML]{E5E5E5}\textbf{100.0\%} \\
\midrule
 & WB\cite{2020_white_blur} & 33.514 & 0.935 & 0.031 & \textbf{100.0\%} & 33.902 & 0.945 & 0.106 & \textbf{98.7\%} & 32.645 & 0.944 & 0.110 & \uline{88.5\%} & 34.662 & 0.958 & 0.047 & \uline{84.0\%} \\
 & SS\cite{2020_white_blur} & 33.500 & 0.959 &0.052 & \textbf{100.0\%} & 34.031 & 0.956 & 0.083 & 83.0\% & 32.882 & 0.960 & 0.080 & 63.0\% & 34.729 & 0.963 & 0.038 & 72.5\% \\
 & AF\cite{ijcai2022p107} & \textbf{53.638} & \textbf{0.999} & \textbf{0.000} & \textbf{100.0\%}& \uline{36.265} & \uline{0.966} & \textbf{0.027} & \textcolor{red}{\textbf{54.6\%}} & 30.693 & 0.924 &0.096& 83.5\% & \textbf{39.245} & \uline{0.983} & \textbf{0.010} & \textcolor{red}{\textbf{17.0\%}} \\
 & SA\cite{2024_saliency_aware} & \uline{37.882} & \uline{0.978} & \uline{0.009} & \uline{99.6\%} &\textbf{36.361} & \textbf{0.970} & \uline{0.049} & 89.9\% & \textbf{34.414} & \textbf{0.964} &\uline{0.064} &84.0\% & \uline{35.275} & 0.964 & \uline{0.037} & 79.5\% \\
 & UA\cite{2024_union-saliency} & 34.451 & 0.931 & 0.110 & \textbf{100.0\%} & 29.817 & 0.865 &0.173 & 69.4\% & 35.693 & 0.964 & \textbf{0.069} & \textcolor{red}{\textbf{44.5\%}} & 35.256 & 0.961 & 0.043 & 45.5\% \\
 & DR\cite{2024_DF_RAP} &35.326 & 0.955 & 0.065 & \uline{99.6\%} & 36.188 & 0.969 & 0.053 &91.5\% & \uline{33.236} & 0.947 & 0.128 & 82.0\% & 35.302 & 0.958 & 0.070 & 77.5\% \\
\multirow{-7}{*}{\begin{tabular}[c]{@{}c@{}}FFHQ\\ \cite{FFHQ}\end{tabular}} & \cellcolor[HTML]{E5E5E5}AEGIS & \cellcolor[HTML]{E5E5E5}31.914 & \cellcolor[HTML]{E5E5E5}0.927 & \cellcolor[HTML]{E5E5E5}0.064 & \cellcolor[HTML]{E5E5E5}\textbf{100.0\%} & \cellcolor[HTML]{E5E5E5}31.997 & \cellcolor[HTML]{E5E5E5}0.935 & \cellcolor[HTML]{E5E5E5}0.125 & \cellcolor[HTML]{E5E5E5}{\uline{93.9\%}} & \cellcolor[HTML]{E5E5E5}{34.141} & \cellcolor[HTML]{E5E5E5}{\uline{0.969}} & \cellcolor[HTML]{E5E5E5}{0.069} & \cellcolor[HTML]{E5E5E5}{\textbf{94.0\%}} & \cellcolor[HTML]{E5E5E5}33.170 & \cellcolor[HTML]{E5E5E5}{\textbf{0.984}} &\cellcolor[HTML]{E5E5E5}0.045 & \cellcolor[HTML]{E5E5E5}\textbf{99.0\%} \\
\midrule
 & WB\cite{2020_white_blur} & 34.376 & 0.938 & 0.036 & \textbf{100.0\%} & 34.053 & 0.936 & 0.148 & \textbf{99.8\%} & 32.828 & 0.932 & 0.143 & \uline{93.0\%} & 35.417 & 0.952 & 0.064 &\uline{98.0\%} \\
 & SS\cite{2020_white_blur} & 34.483 & 0.956 &0.063 & \textbf{100.0\%} & 34.090 & 0.945 & 0.114 & 84.1\% & 32.986 & 0.947 & 0.103 & 77.0\% & 35.349 & 0.952 & 0.063 & 93.5\% \\
 & AF\cite{ijcai2022p107} & \textbf{54.559} & \textbf{0.999} & \textbf{0.000} & \textbf{100.0\%}& \uline{36.826} & \uline{0.967} & \textbf{0.043} & \textcolor{red}{\textbf{55.6\%}} & 31.065 & 0.919 &0.121& 84.0\% & \textbf{39.215} & \uline{0.980} & \textbf{0.016} & 87.5\% \\
 & SA\cite{2024_saliency_aware} & \uline{38.139} & \uline{0.977} & \uline{0.014} & \textbf{100.0\%} &\textbf{37.520} & \textbf{0.978} & \uline{0.048} & 85.5\% & \textbf{35.515} & \textbf{0.971} &\textbf{0.063} &85.0\% & \uline{36.071} & 0.961 & \uline{0.059} & 95.5\% \\
 & UA\cite{2024_union-saliency} & 34.422 & 0.946 & 0.049 & \textbf{100.0\%} & 30.053 & 0.859 &0.249 & 64.5\% & 35.750 & 0.953 & \uline{0.097} &\textcolor{red}{\textbf{53.5\%}} & 35.747 & 0.955 & 0.070 &\textcolor{red}{\textbf{69.5\%}} \\
 & DR\cite{2024_DF_RAP} &35.406 & 0.948 & 0.112 & \uline{99.2\%} & 35.096 & 0.945 & 0.176 &94.9\% & 33.278 & 0.932 & 0.174 & 87.0\% & 35.711 & 0.953 & 0.105 & 96.5\% \\
\multirow{-7}{*}{\begin{tabular}[c]{@{}c@{}}LFW\\ \cite{LFW}\end{tabular}} & \cellcolor[HTML]{E5E5E5}AEGIS & \cellcolor[HTML]{E5E5E5}33.072 & \cellcolor[HTML]{E5E5E5}0.928 & \cellcolor[HTML]{E5E5E5}0.077 & \cellcolor[HTML]{E5E5E5}\textbf{100.0\%} & \cellcolor[HTML]{E5E5E5}31.041 & \cellcolor[HTML]{E5E5E5}0.924 & \cellcolor[HTML]{E5E5E5}0.200 & \cellcolor[HTML]{E5E5E5}{\uline{90.9\%}} & \cellcolor[HTML]{E5E5E5}{\uline{35.500}} & \cellcolor[HTML]{E5E5E5}{\uline{0.972}} & \cellcolor[HTML]{E5E5E5}{0.127} & \cellcolor[HTML]{E5E5E5}{\textbf{98.0\%}} & \cellcolor[HTML]{E5E5E5}34.574 & \cellcolor[HTML]{E5E5E5}{\textbf{0.988}} &\cellcolor[HTML]{E5E5E5}0.066 & \cellcolor[HTML]{E5E5E5}\textbf{100.0\%} \\
\bottomrule
\end{tabular}
\end{table*}

\begin{table*}[t]
\centering
\caption{Robustness evaluation comparison with area under curve (AUC) under image post-processings in white-box settings.}
\label{tab:whitebox_robustness}
\setlength{\tabcolsep}{2.5pt}
\begin{tabular}{c|cccc|c|cccc|c|cccc|c|cccc|c}
\toprule
\multirow{2}{*}{Model} &
\multicolumn{5}{c|}{StarGAN\cite{stargan}} &
\multicolumn{5}{c|}{AttGAN\cite{attgan}} &
\multicolumn{5}{c|}{HiSD\cite{HiSD}} &
\multicolumn{5}{c}{SimSwap\cite{simswap}} \\
\cmidrule(lr){2-21}
 & P1 & P2 & P3 & P4 & Avg. & P1 & P2 & P3 & P4 & Avg. & P1 & P2 & P3 & P4 & Avg. & P1 & P2 & P3 & P4 & Avg. \\
\midrule
WB\cite{2020_white_blur} & 0.168 & \textbf{1.000} & 0.606 & 0.742 & 0.629 & 0.422 & \textbf{0.883} & 0.323 & \textbf{0.679} & \uline{0.577} & 0.214 & \uline{0.665} & 0.323 & \uline{0.522} & \uline{0.431} & \uline{0.917} & \uline{0.918} & 0.827 & \uline{0.894} & \uline{0.889} \\
SS\cite{2020_white_blur} & 0.419 & \textbf{1.000} & \textbf{0.902} & \textbf{0.916} & \textbf{0.809} & 0.287 & 0.603 & \uline{0.339} & 0.538 & 0.442 & 0.161 & 0.458 & \uline{0.530} & 0.449 & 0.400 & 0.845 & 0.822 & \uline{0.914} & 0.893 & 0.869 \\
AF\cite{ijcai2022p107}   & 0.056 & 0.150 & 0.247 & 0.127 & 0.145 & 0.010 & 0.027 & 0.043 & 0.017 & 0.024 & 0.046 & 0.230 & 0.197 & 0.256 & 0.182 & 0.037 & 0.014 & 0.443 & 0.076 & 0.142 \\
SA\cite{2024_saliency_aware}  & 0.204 & 0.907 & 0.457 & 0.625 & 0.548 & 0.308 & 0.666 & 0.239 & 0.500 & 0.428 & 0.234 & 0.531 & 0.300 & 0.437 & 0.376 & 0.883 & 0.851 & 0.817 & 0.859 & 0.853 \\
UA\cite{2024_union-saliency} & 0.130 & 0.882 & 0.459 & 0.568 & 0.510 & 0.107 & 0.217 & 0.098 & 0.161 & 0.146 & 0.068 & 0.245 & 0.173 & 0.208 & 0.173 & 0.536 & 0.542 & 0.647 & 0.557 & 0.571 \\
DR\cite{2024_DF_RAP} & \underline{0.516} & 0.173 & 0.269 & 0.212 & 0.293 & 0.443 & 0.166 & 0.153 & 0.285 & 0.262 & \underline{0.361} & 0.190 & 0.149 & 0.177 & 0.219 & 0.469 & 0.379 & 0.563 & 0.344 & 0.439 \\
\rowcolor[HTML]{E5E5E5} AEGIS & \textbf{0.704} & 0.814 & \underline{0.860} & \underline{0.852} & \underline{0.808} & \textbf{0.623} & \underline{0.709} & \textbf{0.367} & \underline{0.639} & \textbf{0.585} & \textbf{0.901} & \textbf{0.947} & \textbf{0.871} & \textbf{0.906} & \textbf{0.906} & \textbf{0.983} & \textbf{0.954} & \textbf{0.976} & \textbf{0.982} & \textbf{0.974} \\
\bottomrule
\end{tabular}
\end{table*}

\subsubsection{Robustness}
\label{robustness}

\begin{figure*}[]
    \centering
    \includegraphics[width=\textwidth]{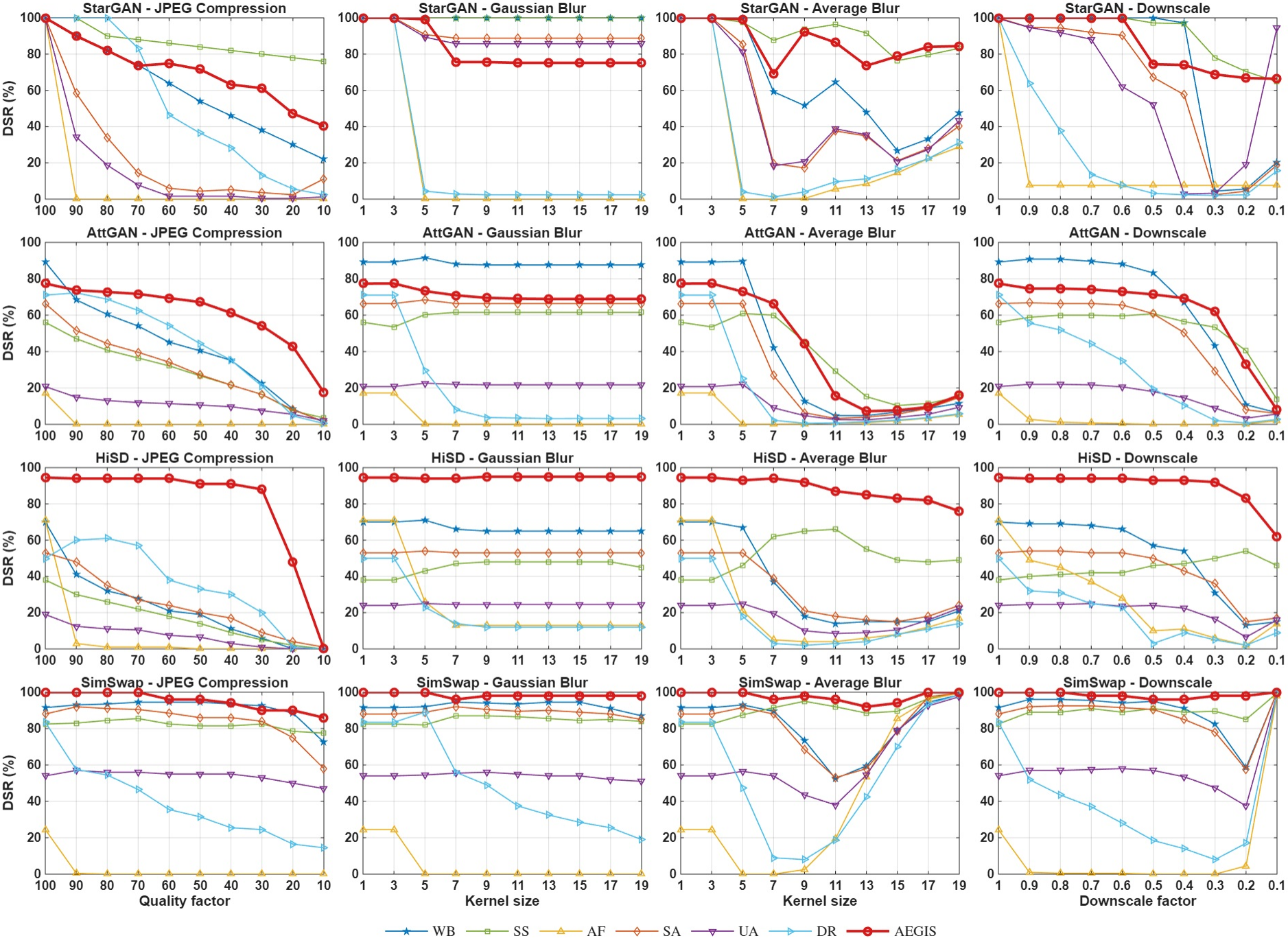}
    \caption{Robustness comparison with SOTA methods in white-box settings }
    \label{robustness_whitebox}
\end{figure*}

Given that social media platforms often apply various image processing operations, we evaluated the robustness of adversarial defenses under common perturbations. Concretely, we post-process adversarial samples using four widely used operations: JPEG compression, Gaussian blur, Average blur, and Downscale. The JPEG quality factor (QF) ranged from 10 to 100, the blur kernel sizes from 1 to 19, and the downscale factor from 0.1 to 1. We then tested the DSR against four manipulation models on the CelebA dataset. Figure~\ref{robustness_whitebox} reports the resulting DSR curves (our method is shown in red). For concise reporting in Table ~\ref{tab:whitebox_robustness}, we denote these four operations as P1–P4 (P1: JPEG; P2: Gaussian blur; P3: average blur; P4: downscaling) and summarize robustness using the area under each curve (AUC).

From Figure~\ref{robustness_whitebox}, AEGIS exhibits consistently strong robustness across all four post-processing operations. It maintains the highest DSR on HiSD and SimSwap over almost all parameter ranges, showing strong resistance to compression- and blur-induced degradations. Although its performance on StarGAN and AttGAN gradually decreases as the processing becomes more aggressive, AEGIS still remains the top or second-best method across most operating points. %demonstrating stable robustness rather than brittle, range-specific gains.

To quantify overall robustness beyond individual curve shapes, we further computed the AUC of each curve and report the results in Table~\ref{tab:whitebox_robustness}. This aggregate measure confirms the same trend: AEGIS achieves the best or near-best average robustness across processing operations, validating that its advantage is not confined to a particular post-processing setting but generalizes across realistic image-degradation pipelines.

\subsection{Black-box Facial Deepfake Scenario}

To further evaluate the extensibility of the proposed method, we conduct black-box experiments where the defense method cannot access the gradients or internal parameters of the deepfake model. Facial forgery models are divided into two categories based on their generation mechanisms: \textit{attribute editing} models (e.g., StarGAN, FPGAN, AttentionGAN) and \textit{face swapping} models (e.g., SimSwap). We assess our method’s robustness on both categories and further perform cross-model transfer attacks to examine its generalization across different deepfake models.

\subsubsection{Defense Effectiveness}

\begin{figure}[]
    \centering
    \includegraphics[width=1\linewidth]{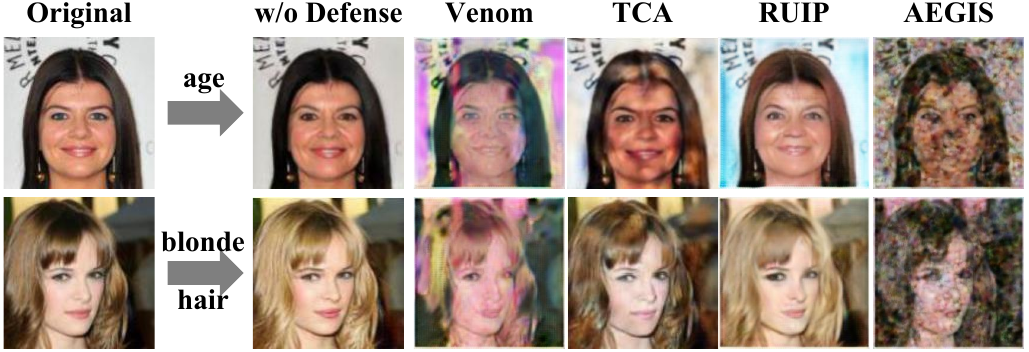}
    \caption{Visualization of the SOTA methods and AEGIS's defense effectiveness against attribute editing in black-box setting.}
    \label{fig:attrediting_blackbox}
\end{figure}

\textit{Attribute editing:}
Table~\ref{tab:blackbox_results_attributediting} and Figure~\ref{fig:attrediting_blackbox} present the quantitative and qualitative results on three representative attribute editing models, that is, StarGAN, FPGAN, and AttentionGAN.

On \textbf{StarGAN}, AEGIS achieves a 41.5\% higher DSR than RUIP. Although AEGIS's DSR (95.5\%) is slightly lower than Venom’s, its protection is substantially more meaningful. Venom achieves a higher DSR, yet its adversarial outputs retain clear identity characteristics, as indicated by its higher ID sim (0.720). This means that the forged faces remain identity-recognizable, thus failing to prevent potential misuse.  
In contrast, AEGIS yields the lowest ID sim (0.111) among all black-box defenses, demonstrating that it suppresses identity information rather than merely disrupting the manipulation process. As visually confirmed in Figure~\ref{fig:attrediting_blackbox}, AEGIS not only prevents successful forgery but also preserves user privacy by avoiding facial stigmatization.
On \textbf{FPGAN}, AEGIS achieves the highest DSR of 94.0\%, about 10\% higher than RUIP. It also attains the best $L_2$ (0.302) and SSIM (0.549), while maintaining the lowest ID sim.
On \textbf{AttentionGAN}, the DSR of AEGIS (68\%) is lower than those of Venom and RUIP. We hypothesize that the lower black-box success rate on AttentionGAN is caused by its localized mask-based editing mechanism, which significantly reduces the transferability of adversarial gradients and suppresses the effect of perturbations outside the attention region. Nevertheless, AEGIS also achieves the strongest disruption of identity features, with the lowest ID sim score of only 0.137, further confirming its effectiveness in safeguarding identity privacy.

\textit{Face swapping:}
We further evaluate the black-box defense performance on the face-swapping model. Note that Venom generates perturbation using a surrogate model designed only for single-image attribute editing and cannot handle the dual-image input required by face swapping; therefore, Venom is excluded from this comparison. We thus compare only RUIP and AEGIS on SimSwap.

\begin{figure}[]
    \centering
    \includegraphics[width=0.9\linewidth]{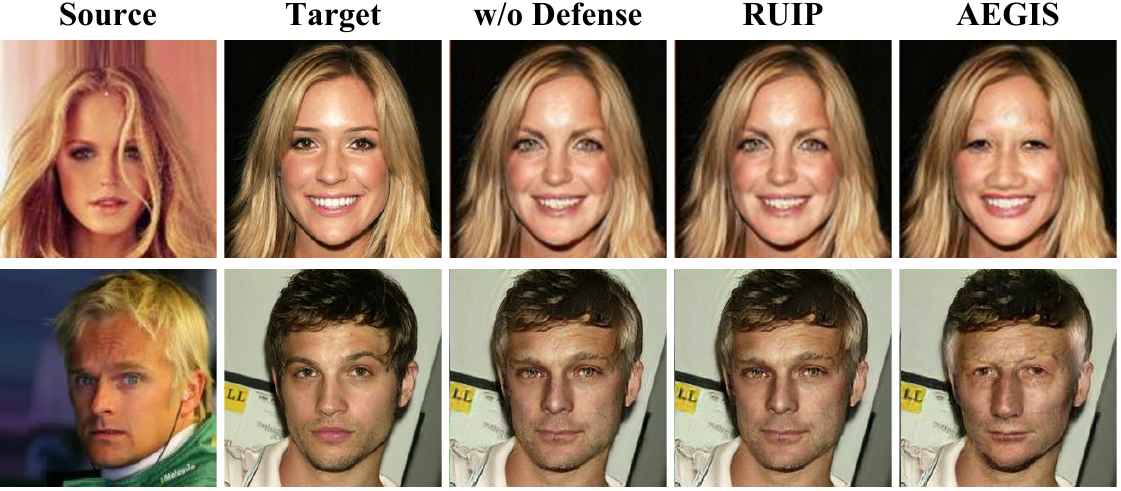}
    \caption{Visualization of the SOTA methods and AEGIS's defense effectiveness against face swapping in black-box setting.}
    \label{fig:faceswap_blackbox}
\end{figure}

As shown in Table~\ref{tab:simswap_blackbox}, RUIP maintains excellent visual fidelity in the generated adversarial faces, yet its DSR is 0, confirming the limitation reported in its original work. In contrast, AEGIS achieves a DSR of 72\%, demonstrating that its gradient estimation strategy can still provide effective protection against face-swapping forgeries.
Figure~\ref{fig:faceswap_blackbox} further illustrates the qualitative results on SimSwap. Compared with the non-defended case and the outputs of RUIP, our method significantly reduces the success of black-box face swapping by preventing the adversary from producing a visually convincing target image that carries the source identity. This yields strong visual defense performance.

\begin{table}[t]
\centering
\caption{Comparison of Black-box Defense Results on Different GAN-based Deepfake Models for Facial Attribute Editing.}
\label{tab:blackbox_results_attributediting}
\setlength{\tabcolsep}{3pt}
\begin{tabular}{c|c|ccc|c}
\hline
\multirow{2}{*}{Models} & \multirow{2}{*}{Method} & \multicolumn{3}{c|}{Defense Effectiveness} & Identity Feature \\ 
\cline{3-6}
& & DSR$\uparrow$ & $L_2$$\uparrow$ & SSIM$\downarrow$ & ID sim$\downarrow$ \\
\hline
\multirow{3}{*}{\begin{tabular}[c]{@{}c@{}}StarGAN\\ \cite{stargan}\end{tabular}} & Venom\cite{huang2021initiative} & \textbf{98.8\%} & 0.157 & 0.699  & 0.720 \\
& RUIP\cite{zhang2025robust}  & 54.0\%& 0.013  & 0.924 & 0.556 \\
& \cellcolor[HTML]{E5E5E5}AEGIS &\cellcolor[HTML]{E5E5E5}95.5\% &\cellcolor[HTML]{E5E5E5}\textbf{0.318} & \cellcolor[HTML]{E5E5E5}\textbf{0.551}  & \cellcolor[HTML]{E5E5E5}\textbf{0.111} \\
\hline
\multirow{3}{*}{\begin{tabular}[c]{@{}c@{}}FPGAN\\ \cite{FPGAN}\end{tabular}}
& Venom\cite{huang2021initiative} & 86.8\% & 0.155 & 0.694 &  0.578 \\
& RUIP\cite{zhang2025robust}  & 84.0\% & 0.011 &  0.932 & 0.844 \\
& \cellcolor[HTML]{E5E5E5}AEGIS & \cellcolor[HTML]{E5E5E5}\textbf{94.0\%} & \cellcolor[HTML]{E5E5E5}\textbf{0.302} & \cellcolor[HTML]{E5E5E5}\textbf{0.549}  & \cellcolor[HTML]{E5E5E5}\textbf{0.043} \\
\hline
\multirow{3}{*}{\begin{tabular}[c]{@{}c@{}}Attention\\ GAN\cite{tang2021attentiongan}\end{tabular}}& Venom\cite{huang2021initiative} & \textbf{100.0\%} & \textbf{0.238}  & \textbf{0.618} & 0.553 \\
& RUIP\cite{zhang2025robust}   &  \textbf{100.0\%} & 0.000 & 0.988 & 0.914 \\
& \cellcolor[HTML]{E5E5E5}AEGIS & \cellcolor[HTML]{E5E5E5}68.0\% & \cellcolor[HTML]{E5E5E5}0.136 & \cellcolor[HTML]{E5E5E5}0.628 & \cellcolor[HTML]{E5E5E5}\textbf{0.137} \\
\hline
\end{tabular}
\end{table}

\begin{table}[t]
\centering
\setlength{\tabcolsep}{3pt}
\caption{Comparison of Defense Effectiveness and Perturbation Imperceptibility between AEGIS and black-box defense method on face-swapping.}
\label{tab:simswap_blackbox}
\begin{tabular}{c|ccc|ccc}
\hline
\multirow{2}{*}{Method} & \multicolumn{3}{c|}{Defense Performance} & \multicolumn{3}{c}{Imperceptibility} \\
\cline{2-7}
 & DSR$\uparrow$ & $L_1$$\downarrow$ & ID sim$\downarrow$ & PSNR$\uparrow$ & SSIM$\uparrow$ & LPIPS$\downarrow$ \\
\hline
RUIP\cite{zhang2025robust} & \textcolor{red}{\textbf{0\%}} & 0.7314 & 0.7136 & \textbf{37.9018} & \textbf{0.9921} & \textbf{0.0007} \\
\rowcolor[HTML]{E5E5E5} AEGIS & \textbf{72\%} & \textbf{0.0472} & \textbf{0.3253} & 31.0390 & 0.8865 & 0.2283 \\
\hline
\end{tabular}
\end{table}

\subsubsection{Imperceptibility}

\begin{figure}[t]
    \centering
    \includegraphics[width=0.9\linewidth]{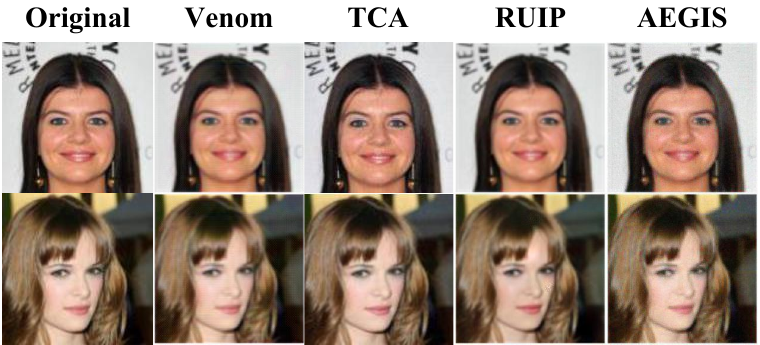}
    \caption{Visualization of perturbation imperceptibility for SOTA methods and AEGIS under the black-box scenario.}
    \label{fig:black_box_imperceptibility}
\end{figure}

In this part, we analyze the trade-off between the visual quality of the adversarial images and defense effectiveness in the black-box scenario. From Table~\ref{tab:blackbox_fidelity} we can see that our method produces slightly lower visual quality metrics compared with other approaches, indicating a modest degradation in pixel- and structure-level fidelity. For AttentionGAN, the DSR of AEGIS is slightly lower than that of the others, possibly due to its attention mechanism, which reinforces key feature regions and reduces the impact of black-box perturbations. Nevertheless, as shown in Fig.~\ref{fig:black_box_imperceptibility}, perturbations remain visually imperceptible to human observers, ensuring that the overall realism of the images is preserved. In exchange for this minor loss in visual quality, the proposed method achieves a substantial gain in DSR, significantly surpassing the values of RUIP and performing comparably to that of Venom. 

\begin{table}[t]
\centering
\caption{Fidelity Comparison of Adversarial Images with SOTA Methods in Black-box Settings.}
\label{tab:blackbox_fidelity}
\setlength{\tabcolsep}{3pt}
\begin{tabular}{c|c|ccc|c}
\hline
\multirow{2}{*}{Models} & \multirow{2}{*}{Method} & \multicolumn{3}{c|}{Imperceptibility} & Defense Effectiveness \\ 
\cline{3-6}
& & PSNR$\uparrow$ & SSIM$\uparrow$ & LPIPS$\downarrow$ & DSR$\uparrow$ \\
\hline

\multirow{3}{*}{\begin{tabular}[c]{@{}c@{}}StarGAN\\ \cite{stargan}\end{tabular}} & Venom\cite{huang2021initiative} & 34.023 & 0.947 & 0.040  & \textbf{98.8\%} \\
& RUIP\cite{zhang2025robust}  & \textbf{35.838} & \textbf{0.981}  & \textbf{0.017} & 54.0\% \\
& \cellcolor[HTML]{E5E5E5}AEGIS & \cellcolor[HTML]{E5E5E5}32.093 & \cellcolor[HTML]{E5E5E5}0.905 & \cellcolor[HTML]{E5E5E5}0.202  & \cellcolor[HTML]{E5E5E5}95.5\% \\
\hline

\multirow{3}{*}{\begin{tabular}[c]{@{}c@{}}FPGAN\\ \cite{FPGAN}\end{tabular}}
& Venom\cite{huang2021initiative} & 34.012 & 0.942 & 0.033 &  86.8\% \\
& RUIP\cite{zhang2025robust}  & \textbf{35.836}  & \textbf{0.978}  & \textbf{0.017} & 84.0\% \\
& \cellcolor[HTML]{E5E5E5}AEGIS & \cellcolor[HTML]{E5E5E5}32.180 & \cellcolor[HTML]{E5E5E5}0.905 & \cellcolor[HTML]{E5E5E5}0.206 &  \cellcolor[HTML]{E5E5E5}\textbf{94.0\%}\\
\hline

\multirow{3}{*}{\begin{tabular}[c]{@{}c@{}}Attention\\ GAN\cite{tang2021attentiongan}\end{tabular}}& Venom\cite{huang2021initiative} & 33.335 & 0.945  & 0.087 & \textbf{100.0\%} \\
& RUIP\cite{zhang2025robust}   & 41.416 & 0.989 & 0.001 & \textbf{100.0\%} \\
& \cellcolor[HTML]{E5E5E5}AEGIS & \cellcolor[HTML]{E5E5E5}30.202& \cellcolor[HTML]{E5E5E5}0.793 & \cellcolor[HTML]{E5E5E5}0.362 & \cellcolor[HTML]{E5E5E5}68.0\% \\
\hline
\end{tabular}
\end{table}

\subsubsection{Transferability}

In the black-box transferability experiments, adversarial samples were first generated on each of the three deepfake models, namely StarGAN, FPGAN, and AttentionGAN, and then tested against the remaining models to evaluate cross-model defense performance.

As shown in Table~\ref{tab:blackbox_trans}, AEGIS consistently achieves higher DSRs than RUIP across all transfer paths, indicating stronger cross-model generalization. When adversarial images are generated on the more complex AttentionGAN and evaluated on other models, AEGIS attains DSRs of 70.4\% on StarGAN and 51.6\% on FPGAN, substantially exceeding those of Venom (43.2\% and 18.4\%, respectively). This suggests that perturbations produced by AEGIS on  attention-based architectures transfer more effectively to heterogeneous manipulation models, reflecting improved cross-model transferability.
We observe a minor exception when StarGAN is used as the source model, where AEGIS yields slightly lower transfer DSRs than Venom. A plausible explanation is that Venom learns a StarGAN-specific perturbation generator, which can better match the inductive biases of similar architectures, but tends to generalize less reliably to models with different structures.

Despite this, AEGIS maintains stable transferability across all transfer settings and avoids catastrophic failures (DSR of 0\%), effectively avoiding the strong model dependency observed in RUIP. In summary, these results demonstrate that AEGIS  provides transferable black-box protection across diverse deepfake models, a capability that existing methods such as Venom and RUIP do not reliably deliver.

\begin{table}[t]
\centering
\caption{Transferability Comparison of Adversarial Images with SOTA Methods in Black-box Settings.}
\label{tab:blackbox_trans}
\setlength{\tabcolsep}{3pt}
\begin{tabular}{c|c|ccc}
\hline
\multirow{2}{*}{Method} & \multirow{2}{*}{Models} & \multicolumn{3}{c}{DSR$\uparrow$} \\ 
\cline{3-5}
& & StarGAN & FPGAN & AttentionGAN  \\
\hline

\multirow{3}{*}{\begin{tabular}[c]{@{}c@{}}Venom\cite{huang2021initiative}\\ \end{tabular}} 
& StarGAN\cite{stargan} & 98.8\% & 91.6\% & 70.0\%   \\
& FPGAN\cite{FPGAN}  & 99.2\% & 86.8\%  & 59.6\%  \\
& AttentionGAN\cite{tang2021attentiongan} & 43.2\% & 18.4\% & 100\%   \\
\hline

\multirow{3}{*}{\begin{tabular}[c]{@{}c@{}}RUIP\cite{zhang2025robust}\\ \end{tabular}}
& StarGAN\cite{stargan} & 54.0\% & 0.0\% & 46.0\%   \\
& FPGAN\cite{FPGAN}  & 0.0\% & 84.0\%  & 24.0\%  \\
& AttentionGAN\cite{tang2021attentiongan} & 0.0\% & 0.0\% & 100\%   \\
\hline

\rowcolor[HTML]{E5E5E5}  & \cellcolor[HTML]{E5E5E5}StarGAN\cite{stargan} & \cellcolor[HTML]{E5E5E5}96.0\% & \cellcolor[HTML]{E5E5E5}30.0\% & \cellcolor[HTML]{E5E5E5}28.0\%   \\
\rowcolor[HTML]{E5E5E5} AEGIS & \cellcolor[HTML]{E5E5E5}FPGAN\cite{FPGAN}  & \cellcolor[HTML]{E5E5E5}86.4\% & \cellcolor[HTML]{E5E5E5}94\%  & \cellcolor[HTML]{E5E5E5}50.4\%  \\
\rowcolor[HTML]{E5E5E5} & \cellcolor[HTML]{E5E5E5}AttentionGAN\cite{tang2021attentiongan} & \cellcolor[HTML]{E5E5E5}70.4\% & \cellcolor[HTML]{E5E5E5}51.6\% & \cellcolor[HTML]{E5E5E5}68.0\%   \\
\hline
\end{tabular}
\end{table}

\subsection{Defense Performance on Diffusion-based Facial Generation Models}

With the rapid advancement of diffusion-based generation, traditional defenses against GAN-based forgeries often fail to generalize to these new architectures. To assess the generalization of our method, we further evaluate it on Arc2Face\cite{papantoniou2024arc2face}, a representative diffusion-based facial deepfake model. Arc2Face couples an ArcFace encoder with a diffusion-based generator to synthesize realistic faces that maintain identity consistency while allowing variation in pose, expression, and background, making it a challenging and realistic benchmark for adversarial defense.

\begin{figure}[]
    \centering
    \includegraphics[width=0.8\linewidth]{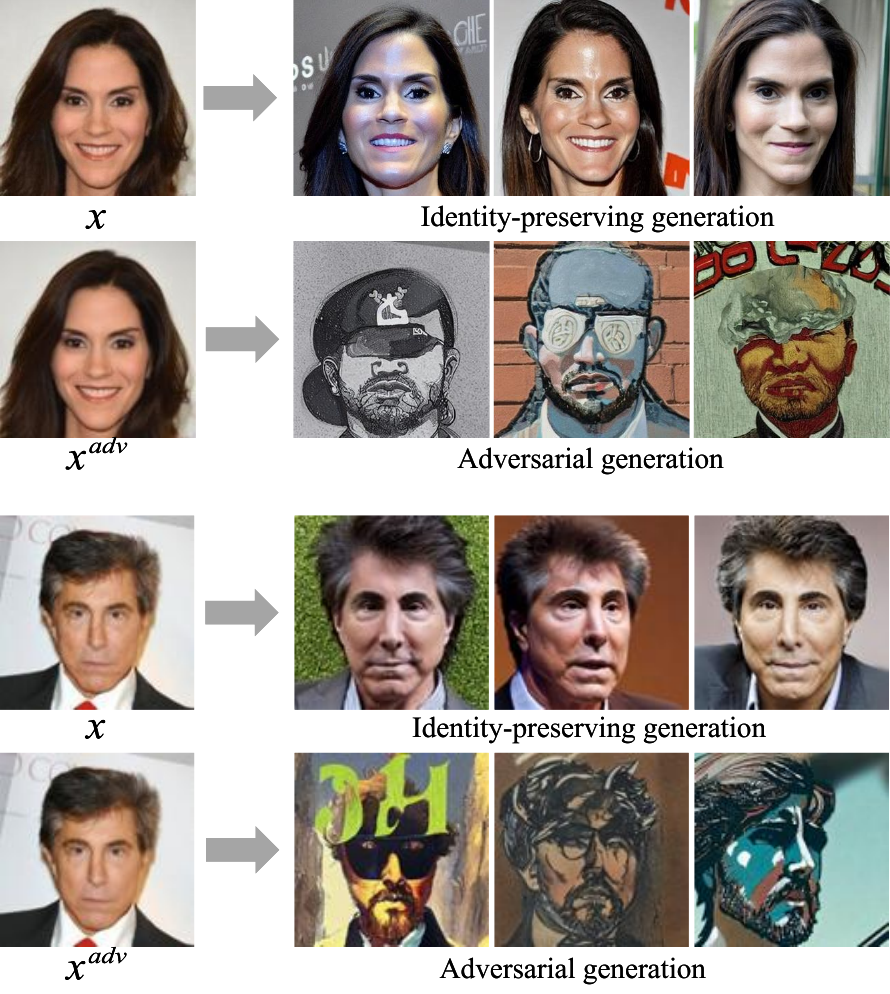}
    \caption{Visualization of defense effectiveness against Arc2Face identity-preserving deepfake model.}
    \label{fig:diffusion-based defense}
\end{figure}

Since both the ArcFace encoder and the diffusion generator are encapsulated and do not expose internal gradients, we conduct the experiment under a \textit{black-box} setting. As shown in Table~\ref{tab:arc2face_defense} and Figure~\ref{fig:diffusion-based defense}, we present the quantitative and qualitative results. AEGIS achieves a DSR of 100\% and an ID Sim of only 0.070, indicating that the generated faces are almost unrecognizable compared with the original faces. The $L_1$ distance reaches 0.266, showing a strong disruption of the forgery process, while the visual quality remains high.
Figure~\ref{fig:diffusion-based defense} provides visual evidence supporting this conclusion. The identity-preserving deepfake results generated by Arc2Face clearly replicate the subject’s facial features under different backgrounds and scenes. In contrast, after applying AEGIS, the generated faces show no trace of the original identity, demonstrating that our method effectively blocks the deepfake synthesis process and achieves complete identity protection.

\begin{table}[t]
\centering
\caption{Results of Imperceptibility and Defense Effectiveness against Arc2Face Facial Generation Model.}
\begin{tabular}{c|ccc}
\hline
\multirow{2}{*}{Imperceptibility} & PSNR$\uparrow$ & SSIM$\uparrow$& LPIPS$\downarrow$ \\ 
\cline{2-4}
&  32.021 &  0.831& 0.115 \\
\hline
\multirow{2}{*}{Defense Effectiveness} & $L_1$ $\uparrow$ & ID sim$\downarrow$ & DSR $\uparrow$\\ 
\cline{2-4}
&  0.266 &  0.070& 100\%\\
\hline
\end{tabular}
\label{tab:arc2face_defense}
\end{table}

\subsection{Ablation Study}

To comprehensively analyze the contribution of each component in the proposed AEGIS framework, we conduct a series of ablation experiments under both post-processed and non–post-processed conditions. Specifically, we investigate four key factors that may influence the defense effectiveness:
\begin{itemize}
    \item The number of diffusion forward steps $T_1$, which controls the strength of latent perturbation (tested at $T_1=20,30,50,100,200,300$);
    \item The number of denoising steps $T_2$ (tested at $T_2=6,10,15,20$).
    \item The injection position of adversarial perturbation during the denoising process, where the perturbation is applied in the last $t$ steps (tested at  $t=1,3,5,7,10$).
    \item The effect of gradient projection on the optimization stability and final defense performance.
\end{itemize}
These experiments jointly evaluate how diffusion dynamics, injection timing, and gradient constraints affect the robustness and visual imperceptibility of the proposed method.

\subsubsection{Impact of Diffusion Forward Step $T_1$}

In the forward process of DDIM, the number of forward steps $T_1$ determines the perturbation intensity applied to the latent representation, which serves as the starting point for the subsequent denoising procedure. The perturbation intensity directly affects the distribution of the adversarial latent variables and consequently impacts both image reconstruction quality and defense effectiveness. We systematically evaluate the performance of AEGIS under different  intensities ($T_1=20, 30, 50, 100, 200, 300$) and report the quantitative results in Table~\ref{tab:forward_noise_intensity}.

Without post-processing, AEGIS maintains a 100\% DSR across all settings, showing that the optimization remains stable. However, as $T_1$ increases, the image quality degrades (PSNR from 32.28 dB to 26.26 dB, SSIM from 0.910 to 0.837), and more importantly, the $L_2$ distance drops from 0.916 to 0.598. This reduction in $L_2$ indicates that the injected latent-space perturbation is attenuated, which directly undermines the adversarial effect of the protected image and lowers the DSR under post-processing conditions.
Under image-processing attacks, this trend becomes more pronounced. A moderate perturbation intensity of $T_1=50$ yields the most stable defense performance, achieving the highest DSRs of 77.2\% and 75.2\% under Gaussian blur and downscaling, respectively, while also preserving the best visual quality as measured by both PSNR and SSIM. 

Overall, the results reveal a clear trade-off between visual fidelity and defense robustness. Considering both perturbation imperceptibility and defense stability under post-processing operations, setting the perturbation injection steps to $T_1=50$ provides the best overall balance. We therefore adopt this configuration as the default in all subsequent experiments.

\begin{table*}[]
\centering
\caption{Ablation Results on the Adversarial Perturbation Intensity in the Forward Process of DDIM}
\vspace{-0.1cm}
\label{tab:forward_noise_intensity}
\begin{tabular}{c|cccc|cc|cc|cc|cc}
\toprule
\multirow{2}{*}{$T_1$} & \multicolumn{4}{c|}{w/o image processing} & \multicolumn{2}{c|}{JPEG compression} & \multicolumn{2}{c|}{Gaussian blur} & \multicolumn{2}{c|}{Average blur} & \multicolumn{2}{c}{Downscale} \\ \cmidrule(l){2-13} 
 & DSR↑ & L2↑ & PSNR↑ & SSIM ↑ & DSR↑ & L2↑ & DSR↑ & L2↑ & DSR↑ & L2↑ & DSR↑ & L2↑ \\ 
\midrule
20 & 100.0\% & 0.916 & 32.282 & 0.910 & 100.0\% & 0.174 & 68.8\% & 0.093 & 72.4\% & 0.097 & 68.8\% & 0.096 \\
30 & 100.0\% & 0.852 & 31.887 & 0.917 & 99.2\% & 0.173 & 76.0\% & 0.123 & 79.8\% & 0.113 & 74.8\% & 0.125 \\
50 & 100.0\% & 0.800 & 32.312 & 0.924 & 95.6\% & 0.155 & 77.2\% & 0.110 & 87.2\% & 0.129 & 75.2\% & 0.111 \\
100 & 100.0\% & 0.585 & 31.903 & 0.923 & 86.4\% & 0.128 & 74.8\% & 0.091 & 89.6\% & 0.126 & 74.0\% & 0.091 \\ 
200 & 100.0\% & 0.577 & 29.254 & 0.891 & 89.2\% & 0.131 & 73.6\% & 0.082 & 84.0\% & 0.096 & 68.4\% & 0.077 \\ 
300 & 100.0\% & 0.598 & 26.263 & 0.837 & 96.4\% & 0.148 & 72.4\% & 0.080 & 82.8\% & 0.101 & 66.4\% & 0.075 \\ \bottomrule
\end{tabular}
\end{table*}

\begin{table*}[]
\centering
\caption{Ablation Results of the Denoising Steps with Adversarial Perturbation Applied at Each step}
\vspace{-0.1cm}
\label{tab:denoising_step}
\begin{tabular}{c|cccc|cc|cc|cc|cc}
\toprule
\multirow{2}{*}{$T_2$} & \multicolumn{4}{c|}{w/o image processing} & \multicolumn{2}{c|}{JPEG Compression} & \multicolumn{2}{c|}{Gaussian Blur} & \multicolumn{2}{c|}{Average Blur} & \multicolumn{2}{c}{Downscales} \\ \cmidrule(l){2-13} 
 & DSR↑ & L2↑ & PSNR↑ & SSIM ↑ & DSR↑ & L2↑ & DSR↑ & L2↑ & DSR↑ & L2↑ & DSR↑ & L2↑ \\ 
\midrule
6 & 100.0\% & 0.688 & 32.920 & 0.936 & 91.2\% & 0.137 & 75.2\% & 0.141 & 88.0\% & 0.123 & 73.2\% & 0.160 \\
10 & 100.0\% & 0.790 & 32.298 & 0.924 & 98.4\% & 0.152 & 77.2\% & 0.110 & 84.8\% & 0.126 & 75.6\% & 0.111 \\
15 & 100.0\% & 0.918 & 31.628 & 0.911 & 98.8\% & 0.175 & 78.0\% & 0.110 & 85.6\% & 0.120 & 76.8\% & 0.112 \\
20 & 100.0\% & 1.042 & 30.812 & 0.893 & 99.2\% & 0.202 & 77.6\% & 0.111 & 87.2\% & 0.126 & 76.0\% & 0.114 \\ \bottomrule
\end{tabular}
\end{table*}

\begin{table*}[]
\centering
\caption{Ablation Results on the Injection Step $t$, where Adversarial Perturbation is Applied from $x_t$ to $x_0$ \\ with 10 Denoising Steps}
\vspace{-0.1cm}
\label{tab:ablation of x_t}
\begin{tabular}{@{}c|cccc|cc|cc|cc|cc@{}}
\toprule
\multirow{2}{*}{$t$} & \multicolumn{4}{c|}{w/o image processing} & \multicolumn{2}{c|}{JPEG Compression} & \multicolumn{2}{c|}{Gaussian Blur} & \multicolumn{2}{c|}{Average Blur} & \multicolumn{2}{c}{Downscales} \\ \cmidrule(l){2-13} 
 & DSR↑ & L2↑ & PSNR↑ & SSIM ↑ & DSR↑ & L2↑ & DSR↑ & L2↑ & DSR↑ & L2↑ & DSR↑ & L2↑ \\ 
\midrule
1 & 100.0\% & 0.259 & 32.206 & 0.958 & 67.2\% & 0.093 & 79.2\% & 0.132 & 66.8\% & 0.079 & 77.2\% & 0.135 \\
3 & 100.0\% & 0.570 & 33.400 & 0.943 & 93.6\% & 0.132 & 74.0\% & 0.118 & 79.6\% & 0.095 & 74.0\% & 0.122 \\
5 & 100.0\% & 0.711 & 33.050 & 0.937 & 90.0\% & 0.132 & 74.4\% & 0.113 & 81.2\% & 0.102 & 71.6\% & 0.115 \\
7 & 100.0\% & 0.748 & 32.749 & 0.932 & 94.4\% & 0.148 & 76.8\% & 0.113 & 86.4\% & 0.121 & 74.0\% & 0.116 \\ 
10 & 100.0\% & 0.790 & 32.298 & 0.924 & 98.4\% & 0.152 & 77.2\% & 0.110 & 84.8\% & 0.126 & 75.6\% & 0.111 \\ 
\bottomrule
\end{tabular}
\end{table*}

\subsubsection{Impact of Denoising Step $T_2$}

The denoising step number $T_2$ determines how many iterations are used to reconstruct the adversarial image from its noisy latent representation. As shown in Table~\ref{tab:denoising_step}, increasing $T_2$ from 6 to 20 strengthens the adversarial perturbation, reflected by the rise of $L_2$ from 0.688 to 1.042 and the corresponding improvement of DSR under JPEG compression from 91.2\% to 99.2\%. This indicates that a larger $T_2$ provides more opportunities for perturbation refinement, allowing the adversarial signal to integrate more deeply into the diffusion trajectory.

However, the stronger perturbations introduced by higher $T_2$ also lead to a gradual degradation of visual quality, with PSNR dropping from 32.92~dB to 30.81~dB and SSIM decreasing from 0.936 to 0.893. The results under different image-processing attacks further confirm this trade-off: moderate denoising depths ($T_2 = 10\sim15$) yield the most balanced defense performance across Gaussian blur, average blur, and downscaling, while overly large $T_2$ values cause quality loss without further robustness gain.

In summary, increasing $T_2$ enhances the embedding strength of perturbations and improves defense performance, but excessive steps harm perceptual quality. Considering both factors, $T_2 = 10$ achieves the best equilibrium between image fidelity and defense effectiveness and is adopted as the default configuration.

\subsubsection{Impact of Perturbation Injection Step $t$}

The parameter $t$ specifies the stage in the denoising process from which adversarial perturbations begin to be injected, continuing from $x_t^*$ down to the final reconstruction $x_0^*$. Table~\ref{tab:ablation of x_t} presents the results under a fixed 10-step denoising schedule. The experiments reveal a consistent trend: starting perturbation injection at later denoising stages enhances defense robustness while maintaining acceptable visual quality.

When injection begins too late, the perturbation affects only the final few iterations, yielding low $L_2$ and poor resilience to common post-processing (DSR below 70\% under JPEG compression). As the injection start moves earlier (increasing $t$), $L_2$ grows and DSRs rise accordingly, with $t=10$ achieving the best defense across attack types while keeping PSNR and SSIM within visually acceptable ranges.

In short, initiating injection from earlier denoising stages (larger $t$) preserves adversarial energy and maximizes the defensive effect; therefore, we adopt $t=10$ by default to balance robustness and image fidelity.

\begin{table}[t]
\centering
\caption{Ablation Results of the DSR With and Without Gradient Projection}
\vspace{-0.1cm}
\label{tab:ablation_gradient}
\setlength{\tabcolsep}{4pt}
\begin{tabular}{c|ccc|ccc}
\toprule
\multirow{2}{*}{Attack} & \multicolumn{3}{c|}{JPEG Compression (QF)} & \multicolumn{3}{c}{Gaussian Blur (Kernel Size)} \\
\cmidrule(lr){2-4} \cmidrule(lr){5-7}
& 90 & 80 & 70 & 1 & 3 & 5 \\
\midrule
w/o GP & 86.8\% & 67.2\% & 65.2\% & 100.0\% & 100.0\% & 66.2\% \\
with GP & 90.0\% & 82.0\% & 73.6\% & 100.0\% & 99.2\% & 75.6\% \\
\midrule
\multirow{2}{*}{Attack} & \multicolumn{3}{c|}{Average Blur (Kernel Size)} & \multicolumn{3}{c}{Downscale (Factor)} \\
\cmidrule(lr){2-4} \cmidrule(lr){5-7}
& 1 & 3 & 5 & 0.7 & 0.6 & 0.5 \\
\midrule
w/o GP & 100.0\% & 100.0\% & 66.4\% & 100.0\% & 99.2\% & 64.8\% \\
with GP & 100.0\% & 99.2\% & 69.2\% & 100.0\% & 100.0\% & 74.4\% \\
\bottomrule
\end{tabular}
\end{table}

\subsubsection{Impact of Gradient Projection}

To resolve gradient conflicts between the defense-oriented $L_{\mathrm{MSE}}$ and the reconstruction-oriented $L_1$ losses, a gradient projection (GP) strategy\cite{GRASP} is introduced to align their update directions during optimization. As shown in Table~\ref{tab:ablation_gradient}, incorporating GP consistently improves DSR under all attack types while preserving image quality.

Compared with the model without GP, the DSR under JPEG compression (QF=70) increases from 65.2\% to 73.6\%, and under downscaling (factor=0.5) rises from 64.8\% to 74.4\%. Similar gains are observed for Gaussian and average blur, confirming that GP stabilizes multi-objective optimization by preventing conflicting gradient updates. In summary, the gradient projection mechanism enables $L_{\mathrm{MSE}}$ and $L_1$ to work synergistically, leading to higher defense robustness without compromising perceptual quality.

\section{Discussion and Conclusion}
AEGIS introduces a new proactive defense paradigm that injects adversarial perturbations into the diffusion denoising trajectory, breaking the conventional $L_\infty$-bounded perturbation design and enabling generalized defense against facial manipulation across GAN- and diffusion-based generators. Extensive experiments demonstrate that AEGIS achieves strong defense effectiveness in both white-box and black-box settings while keeping perturbations visually imperceptible and identity-unrecognizable.

Despite these advantages, AEGIS still has two limitations that warrant further study. First, 
black-box gradient estimation inevitably incurs computational overhead because each update requires multiple forward evaluations. While this overhead is the price paid for model-agnostic deployability, it increases the time to produce protected images and makes the current implementation more suitable for offline publishing or pre-processing workflows than highly latency-sensitive, interactive online scenarios.
Second, AEGIS assumes that the defender can only control the released input image. This assumption aligns with practical proactive protection scenarios, where the adversary typically has access only to publicly shared content rather than the clean source image. Nevertheless, if an adversary can obtain the clean image from alternative sources or apply strong adaptive purification, the defense effectiveness may be weakened.
Future work will explore more efficient and accurate gradient estimation strategies, adaptive perturbation scheduling, query-aware perturbation scheduling, and perceptual-aware refinement to improve both transferability and generation efficiency.

In conclusion, AEGIS represents a step toward practical deepfake defense: it requires no model training, decouples perturbation strength from $L_\infty$ clipping, and operates across manipulation types and threat models. We expect this diffusion-guided defense paradigm to stimulate further research on proactive safeguards against the evolving deepfake ecosystem.

%\section*{Acknowledgments}
%This should be a simple paragraph before the References to thank those individuals and institutions who have supported your work on this article.

\bibliographystyle{IEEEtran}
\bibliography{citation}

\begin{thebibliography}{10}
\providecommand{\url}[1]{#1}
\csname url@samestyle\endcsname
\providecommand{\newblock}{\relax}
\providecommand{\bibinfo}[2]{#2}
\providecommand{\BIBentrySTDinterwordspacing}{\spaceskip=0pt\relax}
\providecommand{\BIBentryALTinterwordstretchfactor}{4}
\providecommand{\BIBentryALTinterwordspacing}{\spaceskip=\fontdimen2\font plus
\BIBentryALTinterwordstretchfactor\fontdimen3\font minus
  \fontdimen4\font\relax}
\providecommand{\BIBforeignlanguage}[2]{{%
\expandafter\ifx\csname l@#1\endcsname\relax
\typeout{** WARNING: IEEEtran.bst: No hyphenation pattern has been}%
\typeout{** loaded for the language `#1'. Using the pattern for}%
\typeout{** the default language instead.}%
\else
\language=\csname l@#1\endcsname
\fi
#2}}
\providecommand{\BIBdecl}{\relax}
\BIBdecl

\bibitem{deepfake_survey}
Y.~Mirsky and W.~Lee, ``The creation and detection of deepfakes: A survey,''
  \emph{ACM computing surveys (CSUR)}, vol.~54, no.~1, pp. 1--41, 2021.

\bibitem{goodfellow2014gan}
I.~J. Goodfellow, J.~Pouget-Abadie, M.~Mirza, B.~Xu, D.~Warde-Farley, S.~Ozair,
  A.~Courville, and Y.~Bengio, ``Generative adversarial nets,'' \emph{Advances
  in neural information processing systems}, vol.~27, 2014.

\bibitem{DDIM}
J.~Song, C.~Meng, and S.~Ermon, ``Denoising diffusion implicit models,'' in
  \emph{International Conference on Learning Representations}, 2021.

\bibitem{DDPM}
J.~Ho, A.~Jain, and P.~Abbeel, ``Denoising diffusion probabilistic models,''
  \emph{Advances in neural information processing systems}, vol.~33, pp.
  6840--6851, 2020.

\bibitem{maras2019determining}
M.-H. Maras and A.~Alexandrou, ``Determining authenticity of video evidence in
  the age of artificial intelligence and in the wake of deepfake videos,''
  \emph{The international journal of evidence \& proof}, vol.~23, no.~3, pp.
  255--262, 2019.

\bibitem{ajder2019state}
H.~Ajder, G.~Patrini, F.~Cavalli, and L.~Cullen, ``The state of deepfakes:
  Landscape, threats, and impact,'' \emph{Amsterdam: Deeptrace}, vol.~27, 2019.

\bibitem{westerlund2019emergence}
M.~Westerlund, ``The emergence of deepfake technology: A review,''
  \emph{Technology innovation management review}, vol.~9, no.~11, 2019.

\bibitem{zhu2025hiding}
D.~Zhu, Y.~Li, B.~Wu, J.~Zhou, Z.~Wang, and S.~Lyu, ``Hiding faces in plain
  sight: Defending deepfakes by disrupting face detection,'' \emph{IEEE
  Transactions on Dependable and Secure Computing}, 2025.

\bibitem{wang2024deepfake}
T.~Wang, X.~Liao, K.~P. Chow, X.~Lin, and Y.~Wang, ``Deepfake detection: A
  comprehensive survey from the reliability perspective,'' \emph{ACM Computing
  Surveys}, vol.~57, no.~3, pp. 1--35, 2024.

\bibitem{lu2023detection}
W.~Lu, L.~Liu, B.~Zhang, J.~Luo, X.~Zhao, Y.~Zhou, and J.~Huang, ``Detection of
  deepfake videos using long-distance attention,'' \emph{IEEE transactions on
  neural networks and learning systems}, vol.~35, no.~7, pp. 9366--9379, 2023.

\bibitem{yin2023dynamic}
Q.~Yin, W.~Lu, B.~Li, and J.~Huang, ``Dynamic difference learning with
  spatio--temporal correlation for deepfake video detection,'' \emph{IEEE
  Transactions on Information Forensics and Security}, vol.~18, pp. 4046--4058,
  2023.

\bibitem{wang2025idcnet}
Z.~Wang, Y.~Chen, Y.~Yao, M.~Han, W.~Xing, and M.~Li, ``Idcnet: Image
  decomposition and cross-view distillation for generalizable deepfake
  detection,'' \emph{IEEE Transactions on Information Forensics and Security},
  2025.

\bibitem{deng2024towards}
J.~Deng, C.~Lin, P.~Hu, C.~Shen, Q.~Wang, Q.~Li, and Q.~Li, ``Towards
  benchmarking and evaluating deepfake detection,'' \emph{IEEE Transactions on
  Dependable and Secure Computing}, vol.~21, no.~6, pp. 5112--5127, 2024.

\bibitem{zhang2024deepfake}
B.~Zhang, Q.~Yin, W.~Lu, and X.~Luo, ``Deepfake detection and localization
  using multi-view inconsistency measurement,'' \emph{IEEE Transactions on
  Dependable and Secure Computing}, 2024.

\bibitem{mittal2020emotions}
T.~Mittal, U.~Bhattacharya, R.~Chandra, A.~Bera, and D.~Manocha, ``Emotions
  don't lie: An audio-visual deepfake detection method using affective cues,''
  in \emph{Proceedings of the 28th ACM international conference on multimedia},
  2020, pp. 2823--2832.

\bibitem{Lin_2024_CVPR}
L.~Lin, X.~He, Y.~Ju, X.~Wang, F.~Ding, and S.~Hu, ``Preserving fairness
  generalization in deepfake detection,'' in \emph{Proceedings of the IEEE/CVF
  Conference on Computer Vision and Pattern Recognition (CVPR)}, June 2024, pp.
  16\,815--16\,825.

\bibitem{carlini2020evading}
N.~Carlini and H.~Farid, ``Evading deepfake-image detectors with white-and
  black-box attacks,'' in \emph{Proceedings of the IEEE/CVF conference on
  computer vision and pattern recognition workshops}, 2020, pp. 658--659.

\bibitem{hussain2021adversarial}
S.~Hussain, P.~Neekhara, M.~Jere, F.~Koushanfar, and J.~McAuley, ``Adversarial
  deepfakes: Evaluating vulnerability of deepfake detectors to adversarial
  examples,'' in \emph{Proceedings of the IEEE/CVF winter conference on
  applications of computer vision}, 2021, pp. 3348--3357.

\bibitem{meng2024ava}
X.~Meng, L.~Wang, S.~Guo, L.~Ju, and Q.~Zhao, ``Ava: Inconspicuous attribute
  variation-based adversarial attack bypassing deepfake detection,'' in
  \emph{2024 IEEE Symposium on Security and Privacy (SP)}.\hskip 1em plus 0.5em
  minus 0.4em\relax IEEE, 2024, pp. 74--90.

\bibitem{2020_white_blur}
N.~Ruiz, S.~A. Bargal, and S.~Sclaroff, ``Disrupting deepfakes: adversarial
  attacks against conditional image translation networks and facial
  manipulation systems,'' in \emph{Proc. of European Conference on Computer
  Vision (ECCV) Workshops}, 2020, pp. 236--251.

\bibitem{ijcai2022p107}
R.~Wang, Z.~Huang, Z.~Chen, L.~Liu, J.~Chen, and L.~Wang, ``Anti-forgery:
  Towards a stealthy and robust deepfake disruption attack via adversarial
  perceptual-aware perturbations,'' in \emph{Proceedings of the Thirty-First
  International Joint Conference on Artificial Intelligence, {IJCAI-22}}, 2022,
  pp. 761--767.

\bibitem{2024_saliency_aware}
Q.~Li, M.~Gao, G.~Zhang, and W.~Zhai, ``Defending deepfakes by saliency-aware
  attack,'' \emph{IEEE Transactions on Computational Social Systems}, vol.~11,
  no.~4, pp. 5060--5067, 2024.

\bibitem{2024_union-saliency}
G.~Zhang, M.~Gao, Q.~Li, W.~Zhai, G.~Zou, and G.~Jeon, ``Disrupting deepfakes
  via union-saliency adversarial attack,'' \emph{IEEE Transactions on Consumer
  Electronics}, vol.~70, no.~1, pp. 2018--2026, 2024.

\bibitem{2024_DF_RAP}
Z.~Qu, Z.~Xi, W.~Lu, X.~Luo, Q.~Wang, and B.~Li, ``Df-rap: A robust adversarial
  perturbation for defending against deepfakes in real-world social network
  scenarios,'' \emph{IEEE Transactions on Information Forensics and Security},
  vol.~19, pp. 3943--3957, 2024.

\bibitem{huang2021initiative}
Q.~Huang, J.~Zhang, W.~Zhou, W.~Zhang, and N.~Yu, ``Initiative defense against
  facial manipulation,'' in \emph{Proceedings of the AAAI conference on
  artificial intelligence}, vol.~35, no.~2, 2021, pp. 1619--1627.

\bibitem{2023_TIFS_Black_box}
J.~Dong, Y.~Wang, J.~Lai, and X.~Xie, ``Restricted black-box adversarial attack
  against deepfake face swapping,'' \emph{IEEE Transactions on Information
  Forensics and Security}, vol.~18, pp. 2596--2608, 2023.

\bibitem{zhang2025robust}
Y.~Zhang, W.~Lin, Z.~Tian, G.~Min, J.~Xu, and Y.~Xu, ``Robust and unstigmatized
  imperceptible perturbations for rendering face manipulation ineffective,''
  \emph{IEEE Transactions on Information Forensics and Security}, 2025.

\bibitem{stargan}
Y.~Choi, M.~Choi, M.~Kim, J.-W. Ha, S.~Kim, and J.~Choo, ``Stargan: Unified
  generative adversarial networks for multi-domain image-to-image
  translation,'' in \emph{Proceedings of The IEEE/CVF Conference on Computer
  Vision and Pattern Recognition (CVPR)}, 2018, pp. 8789--8797.

\bibitem{attgan}
Z.~He, W.~Zuo, M.~Kan, S.~Shan, and X.~Chen, ``Attgan: facial attribute editing
  by only changing what you want,'' \emph{IEEE Transactions on Image
  Processing}, vol.~28, no.~11, pp. 5464--5478, 2019.

\bibitem{HiSD}
X.~Li, S.~Zhang, J.~Hu, L.~Cao, X.~Hong, X.~Mao, F.~Huang, Y.~Wu, and R.~Ji,
  ``Image-to-image translation via hierarchical style disentanglement,'' in
  \emph{Proceedings of the IEEE/CVF Conference on Computer Vision and Pattern
  Recognition (CVPR)}, 2021, pp. 8635--8644.

\bibitem{FPGAN}
M.~M.~R. Siddiquee, Z.~Zhou, N.~Tajbakhsh, R.~Feng, M.~B. Gotway, Y.~Bengio,
  and J.~Liang, ``Learning fixed points in generative adversarial networks:
  From image-to-image translation to disease detection and localization,'' in
  \emph{Proceedings of the IEEE/CVF international conference on computer
  vision}, 2019, pp. 191--200.

\bibitem{tang2021attentiongan}
H.~Tang, H.~Liu, D.~Xu, P.~H. Torr, and N.~Sebe, ``Attentiongan: Unpaired
  image-to-image translation using attention-guided generative adversarial
  networks,'' \emph{IEEE transactions on neural networks and learning systems},
  vol.~34, no.~4, pp. 1972--1987, 2021.

\bibitem{simswap}
R.~Chen, X.~Chen, B.~Ni, and Y.~Ge, ``Simswap: an efficient framework for high
  fidelity face swapping,'' in \emph{Proceedings of The 28th ACM International
  Conference on Multimedia}, 2020, pp. 2003--2011.

\bibitem{li2020advancing}
L.~Li, J.~Bao, H.~Yang, D.~Chen, and F.~Wen, ``Advancing high fidelity identity
  swapping for forgery detection,'' in \emph{Proceedings of the IEEE/CVF
  conference on computer vision and pattern recognition}, 2020, pp. 5074--5083.

\bibitem{papantoniou2024arc2face}
F.~P. Papantoniou, A.~Lattas, S.~Moschoglou, J.~Deng, B.~Kainz, and
  S.~Zafeiriou, ``Arc2face: A foundation model for id-consistent human faces,''
  in \emph{European Conference on Computer Vision}.\hskip 1em plus 0.5em minus
  0.4em\relax Springer, 2024, pp. 241--261.

\bibitem{attack_as_defense}
C.-Y. Yeh, H.-W. Chen, H.-H. Shuai, D.-N. Yang, and M.-S. Chen, ``Attack as the
  best defense: nullifying image-to-image translation gans via limit-aware
  adversarial attack,'' in \emph{Proceedings of the IEEE/CVF International
  Conference on Computer Vision (ICCV)}, 2021, pp. 16\,168--16\,177.

\bibitem{Huang2021CMUAWatermarkAC}
H.~Huang, Y.~Wang, Z.~Chen, Y.~Zhang, Y.~Li, Z.~Tang, W.~Chu, J.~Chen, W.~Lin,
  and K.-K. Ma, ``Cmua-watermark: a cross-model universal adversarial watermark
  for combating deepfakes,'' in \emph{Proceedings of The AAAI Conference on
  Artificial Intelligence}, vol.~36, no.~1, 2022, pp. 989--997.

\bibitem{GRASP}
Y.~Li, L.~Xue, D.~Lin, Q.~Li, H.~Tian, and H.~Wang, ``Towards imperceptible
  adversarial defense: A gradient-driven shield against facial manipulations,''
  \emph{arXiv preprint arXiv:2510.01699}, 2025.

\bibitem{ilyas2018black}
A.~Ilyas, L.~Engstrom, A.~Athalye, and J.~Lin, ``Black-box adversarial attacks
  with limited queries and information,'' in \emph{International conference on
  machine learning}.\hskip 1em plus 0.5em minus 0.4em\relax PMLR, 2018, pp.
  2137--2146.

\bibitem{celeba}
Z.~Liu, P.~Luo, X.~Wang, and X.~Tang, ``Deep learning face attributes in the
  wild,'' in \emph{Proceedings of the IEEE/CVF International Conference on
  Computer Vision (ICCV)}, 2015, pp. 3730--3738.

\bibitem{FFHQ}
T.~Karras, S.~Laine, and T.~Aila, ``A style-based generator architecture for
  generative adversarial networks,'' in \emph{Proceedings of The IEEE/CVF
  Conference on Computer Vision and Pattern Recognition (CVPR)}, 2019, pp.
  4401--4410.

\bibitem{LFW}
G.~B. Huang, M.~Mattar, T.~Berg, and E.~Learned-Miller, ``Labeled faces in the
  wild: a database forstudying face recognition in unconstrained
  environments,'' in \emph{Workshop on Faces in'{R}eal-{L}ife'Images:
  Detection, Alignment, and Recognition}, 2008.

\bibitem{arcface}
J.~Deng, J.~Guo, J.~Yang, N.~Xue, I.~Kotsia, and S.~Zafeiriou, ``Arcface:
  Additive angular margin loss for deep face recognition,'' \emph{IEEE
  Transactions on Pattern Analysis and Machine Intelligence}, vol.~44, no.~10,
  pp. 5962--5979, 2022.

\bibitem{LPIPS}
R.~Zhang, P.~Isola, A.~A. Efros, E.~Shechtman, and O.~Wang, ``The unreasonable
  effectiveness of deep features as a perceptual metric,'' in \emph{Proceedings
  of the IEEE conference on computer vision and pattern recognition}, 2018, pp.
  586--595.

\end{thebibliography}

%If you have an EPS/PDF photo (graphicx package needed), extra braces are needed around the contents of the optional argument to biography to prevent the LaTeX parser from getting confused when it sees the complicated $\backslash${\tt{includegraphics}} command within an optional argument. (You can create your own custom macro containing the $\backslash${\tt{includegraphics}} command to make things simpler here.)
 

% \bf{If you include a photo:}
\vspace{-1.5cm}
\begin{IEEEbiography}[{\includegraphics[width=1in,height=1.25in,clip,keepaspectratio]{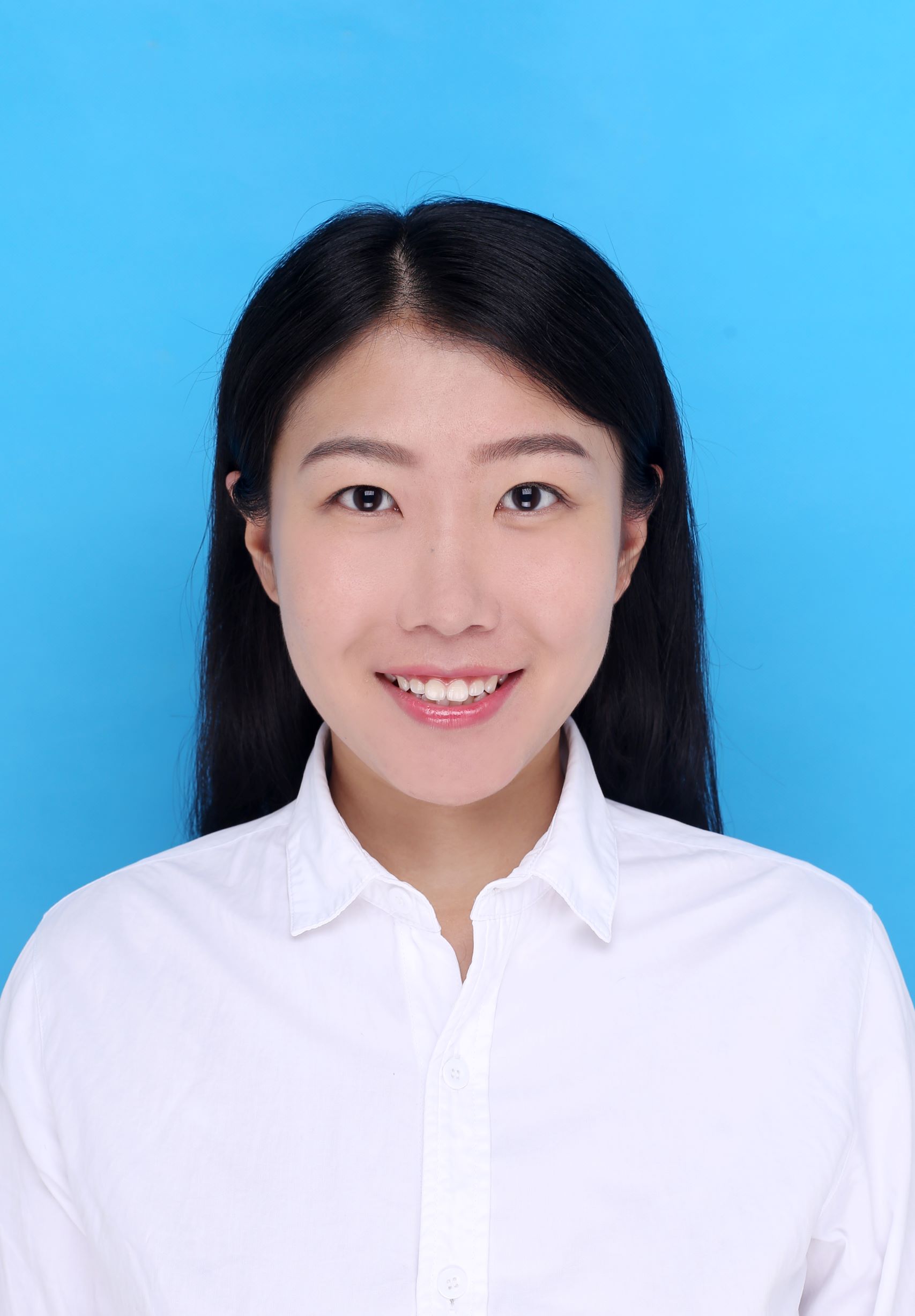}}]{Yue Li}
is a lecturer in the College of Computer Science and Technology at Huaqiao University. She received her B.S. and Ph.D. degrees from Southwest Jiaotong University, Chengdu, China, in 2015 and 2022, respectively. From 2019 to 2021, she was a visiting Ph.D. student at the University of Siena, Italy, under the supervision of Prof. Mauro Barni. Her research interests include multimedia information security, information hiding, deep neural network watermarking, and AI security.
\end{IEEEbiography}
\vspace{-1.3cm}
\begin{IEEEbiography}
[{\includegraphics[width=1in,height=1.25in,clip,keepaspectratio]{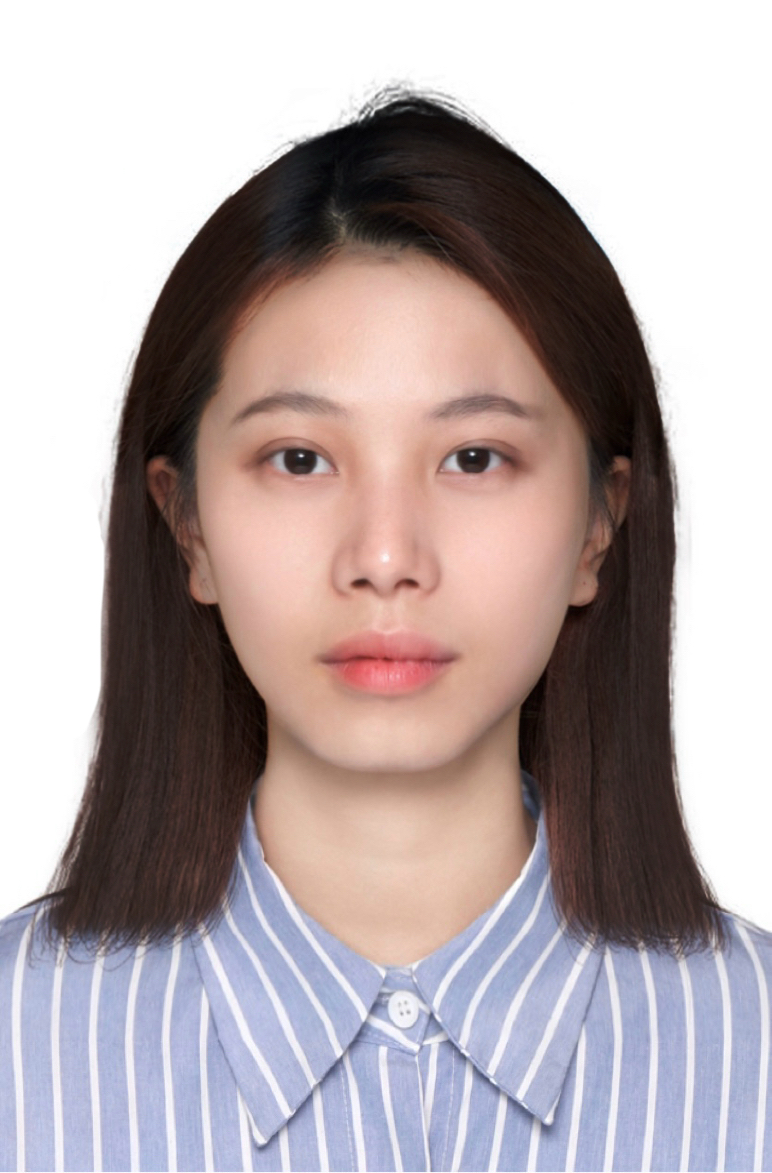}}]{Linying Xue} is currently pursuing her M.S. degree at the College of Computer Science and Technology, National Huaqiao University, Xiamen, China. Her research interests include multimedia information security and AI security.
\end{IEEEbiography}
\vspace{-1.3cm}
\begin{IEEEbiography}
[{\includegraphics[width=1in,height=1.25in,clip,keepaspectratio]{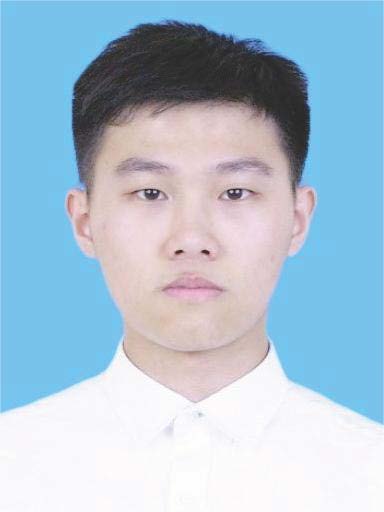}}]{Kaiqing Lin} received the B.S. degree from Shenzhen University, Shenzhen, China, in 2021. He is currently pursuing the Ph.D. degree in Information and Communication Engineering at the College of Electronic and Information Engineering, Shenzhen University, Shenzhen, China. His research interests include multimedia forensics and image steganalysis.
\end{IEEEbiography}
\vspace{-1cm}
\begin{IEEEbiography}
[{\includegraphics[width=1in,height=1.25in,clip,keepaspectratio]{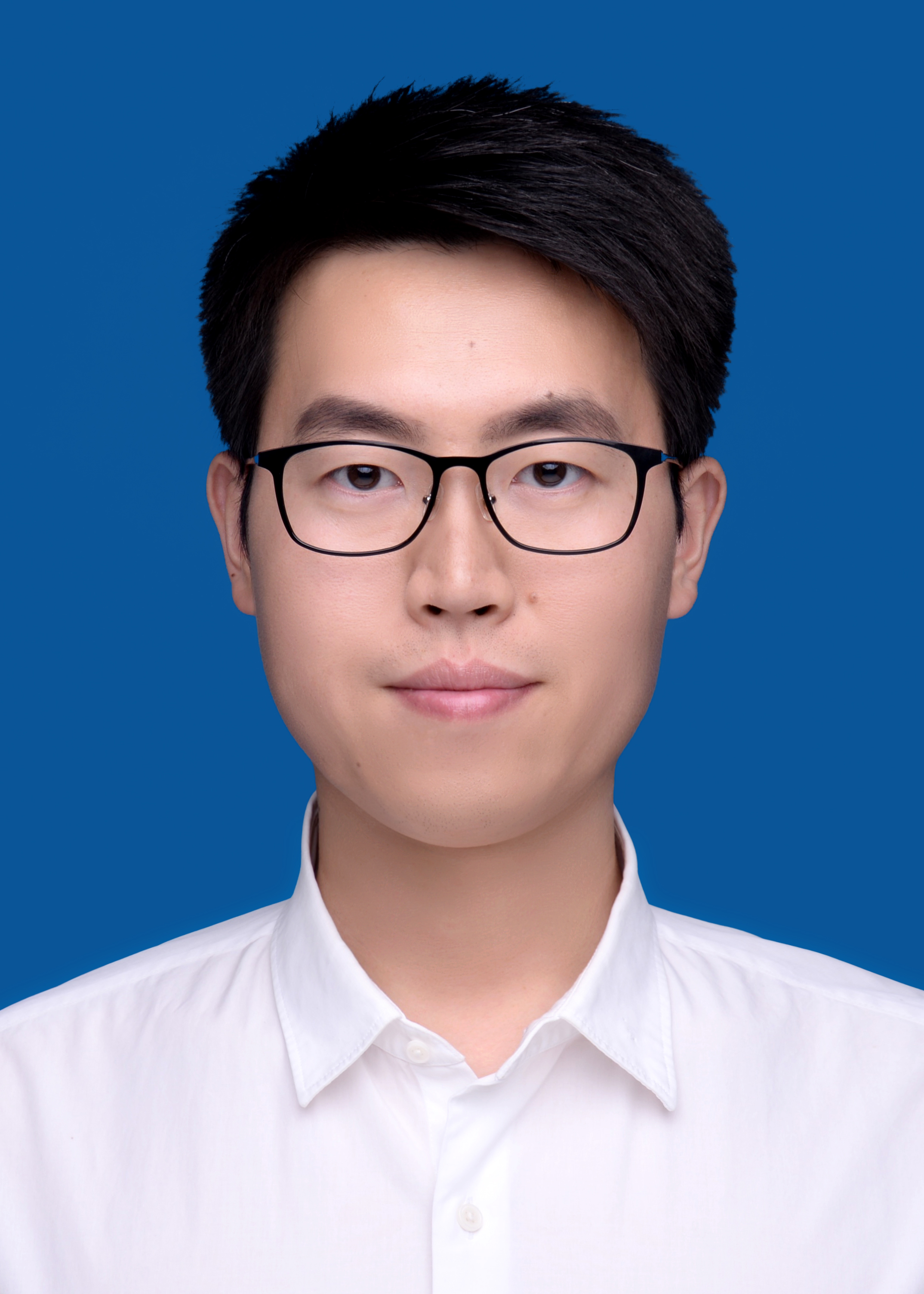}}]{Hanyu Quan} is currently a lecturer in the College of Computer Science and Technology at Huaqiao University, China. He received his PhD degree in Information Security from Xidian University at 2019. He was a visiting PhD student in the Dept. of ECE at the University of Arizona from 2015 to 2017. His research interests include data security and privacy, AI security, and applied cryptography. He is a senior member of IEEE.
\end{IEEEbiography}
\vspace{-1cm}
\begin{IEEEbiography}
[{\includegraphics[width=1in,height=1.25in,clip,keepaspectratio]{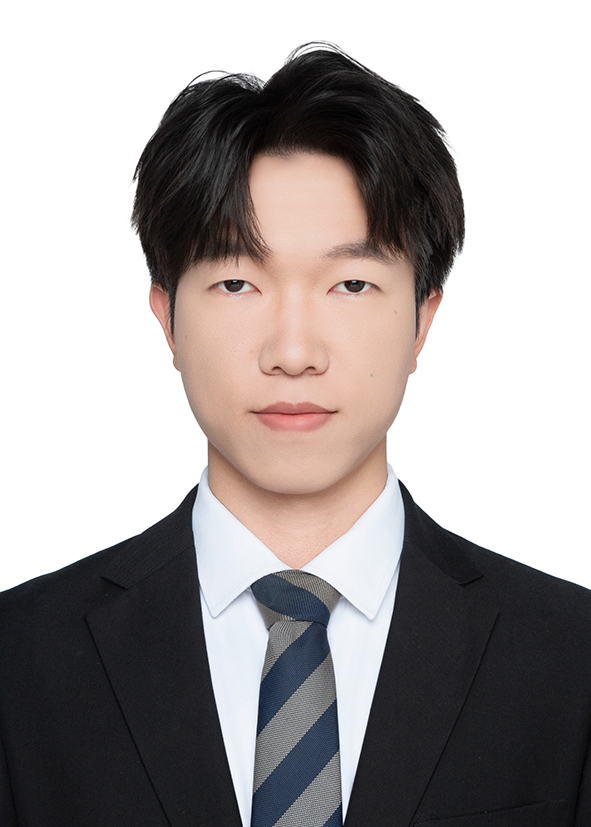}}]{Dongdong Lin} is currently a lecturer with the College of Computer Science and Technology in Huaqiao University. He received his M.Sc. in Computer Science in 2015 from Xiangtan University. He got his Ph.D. in Informatics and Communications in 2025 from Shenzhen University. He was a visiting student with the VIPP lab from 2020 to 2022 at the University of Siena, under the supervision of Prof. Mauro Barni. He has been studying the application of image processing techniques to copyright protection and authentication of multimedia. 
\end{IEEEbiography}
\vspace{-1cm}
\begin{IEEEbiography}
[{\includegraphics[width=1in,height=1.25in,clip,keepaspectratio]{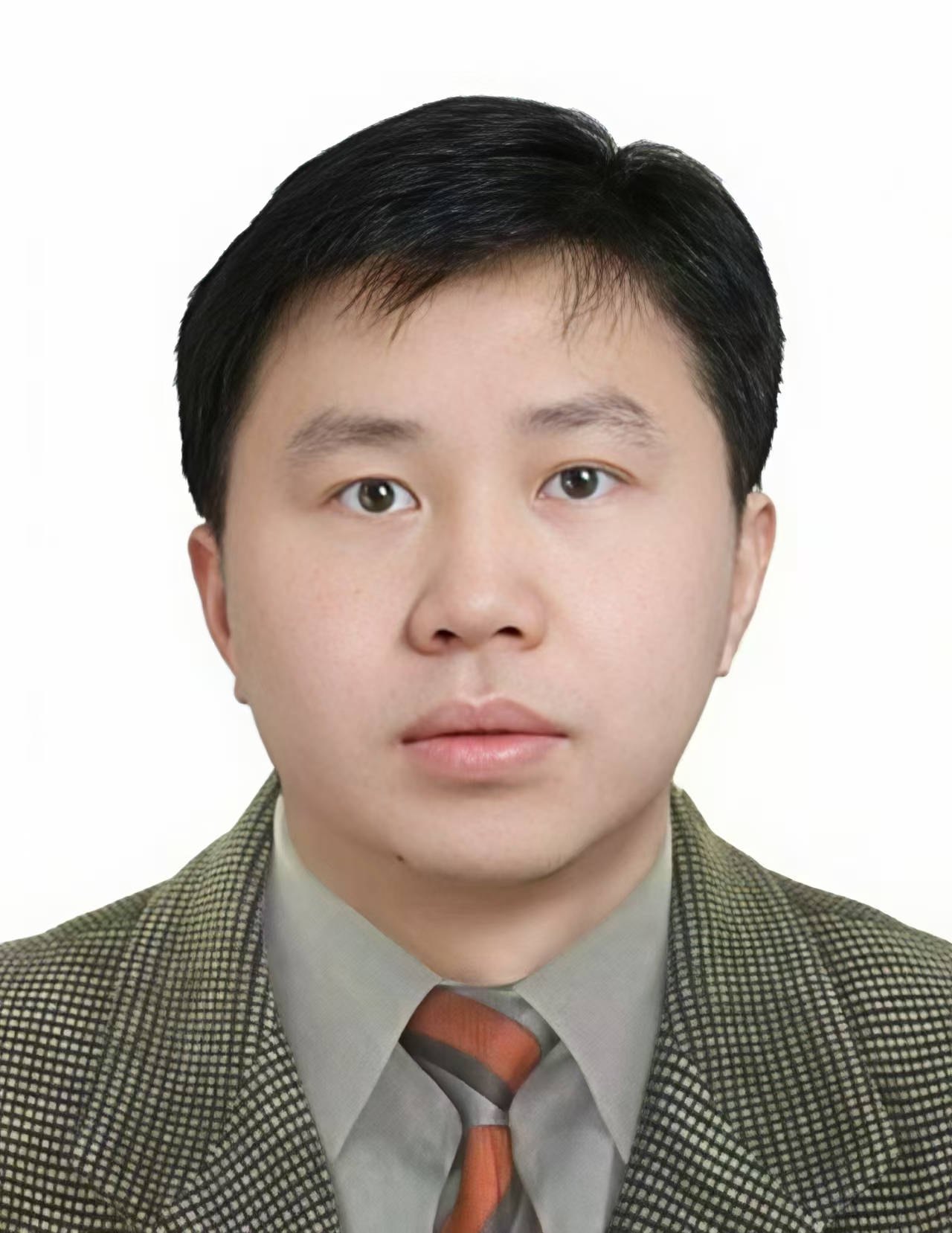}}]{Hui Tian} received the Ph.D. degree in computer science and technology from Huazhong University of Science and Technology, Wuhan, China, in 2010. He is currently a Professor with the College of Computer Science and Technology, Huaqiao University, Xiamen, China. He has published more than 110 papers in refereed journals, proceedings of conferences, and books. His current research interests include network and information security, cloud security, AI security, multimedia security, and digital forensics.
\end{IEEEbiography}
\vspace{-1cm}
\begin{IEEEbiography}
[{\includegraphics[width=1in,height=1.25in,clip,keepaspectratio]{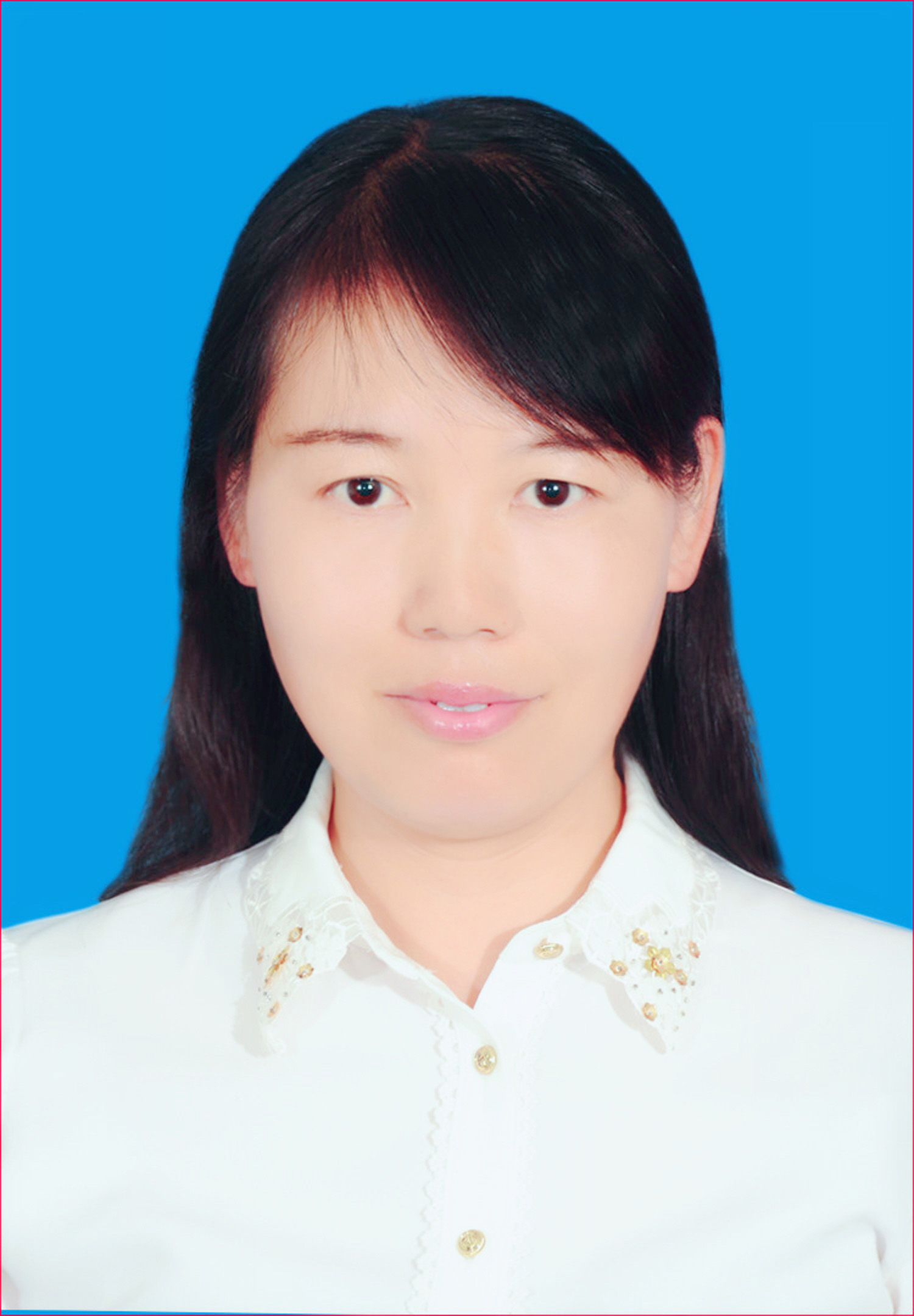}}]{Hongxia Wang} is a professor in the School of Cyber Science and Engineering at Sichuan University. She received her M.S. and Ph.D. degrees from University of Electronic Science and Technology of China, Chengdu, in 1999 and 2002, respectively. She pursued postdoctoral research work at Shanghai Jiao Tong University from 2002 to 2004. She was a professor at Southwest Jiaotong University from 2004 to 2018. Her research interests include multimedia information security, information hiding, digital watermarking, digital forensics, and intelligent information processing. She has published over 200 peer-reviewed research papers and won 18 authorized patents.
\end{IEEEbiography}
\vspace{-1cm}
\begin{IEEEbiography}
[{\includegraphics[width=1in,height=1.25in,clip,keepaspectratio]{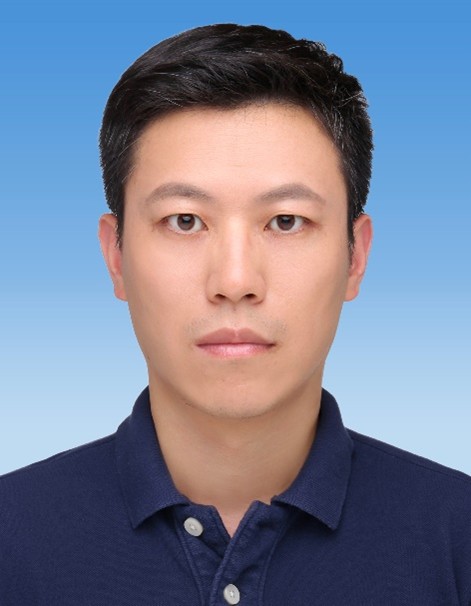}}]{Bin Wang} is a professor of Zhejiang Key Laboratory of Artificial Intelligence of Things (AIoT) Network and Data Security and Xidian University. He received the Ph.D. degree from the China National Digital Switching System Engineering \& Technological R\& D Center. His research interests mainly include Internet of Things security, cryptography, artificial intelligence security, new network security architecture and so on.
\end{IEEEbiography}



\vfill

\end{document}